\documentclass[oldversion]{aa}
\usepackage{txfonts}
\usepackage{graphicx}
\usepackage{epstopdf}
\usepackage{natbib}
\usepackage{rotating}
\bibpunct{(}{)}{;}{a}{}{,}

\newcommand{\rpms}[1]{{#1}\,\mathrm{rad\,m}^{-2}}

\begin{document}

\title{The Westerbork SINGS Survey II.}
\subtitle{Polarization, Faraday Rotation, and Magnetic Fields}

\author{G. Heald\inst{1}
   \and R. Braun\inst{2}
   \and R. Edmonds\inst{3}}

\offprints{George Heald, \email{heald@astron.nl}}

\institute{Netherlands Institute for Radio Astronomy (ASTRON), Postbus 2, 7990 AA Dwingeloo, The Netherlands
      \and CSIRO-ATNF, PO Box 76, Epping, NSW 1710, Australia
      \and New Mexico State University, Department of Astronomy, PO Box 30001, MSC 4500, Las Cruces, New Mexico, 88003-8001, USA}

\date{Received 1 April 2009 / Accepted 20 May 2009}

\abstract{A sample of large northern Spitzer Infrared Nearby Galaxies
  Survey (SINGS) galaxies has recently been observed with the
  Westerbork Synthesis Radio Telescope (WSRT). We present observations
  of the linearly polarized radio continuum emission in this
  WSRT-SINGS galaxy sample. Of the 28 galaxies treated in this paper,
  21 are detected in polarized radio continuum at 18- and 22-cm
  wavelengths. We utilize the rotation measure synthesis
  (RM-Synthesis) method, as implemented by Brentjens \& de Bruyn, to
  coherently detect polarized emission from a large fractional
  bandwidth, while simultaneously assessing the degree of Faraday
  rotation experienced by the radiation along each line-of-sight. This
  represents the first time that the polarized emission and its
  Faraday rotation have been systematically probed down to
  $\sim10~\mu$Jy~beam$^{-1}$ RMS for a large sample of
  galaxies. Non-zero Faraday rotation is found to be ubiquitous in all
  of the target fields, from both the Galactic foreground and the
  target galaxies themselves. In this paper, we present an overview of
  the polarized emission detected in each of the WSRT-SINGS
  galaxies. The most prominent trend is a systematic modulation of the
  polarized intensity with galactic azimuth, such that a global
  minimum in the polarized intensity is seen toward the kinematically
  receding major axis. The implied large-scale magnetic field geometry
  is discussed in a companion paper. A second novel result is the
  detection of multiple nuclear Faraday depth components that are
  offset to both positive and negative RM by $\rpms{100-200}$ in all
  targets that host polarized (circum-)nuclear emission.}

\maketitle

\section{Introduction\label{section:intro}}

In the study of star formation properties and evolution of galaxies,
an important ingredient is the magnetic field content of the ISM. Yet
the precise role that magnetic fields play in regulating star
formation, and the role of magnetic fields in the evolution of galaxy
disks, is still far from well understood. One reason for this gap is
that a systematic survey of magnetic field content in galaxies over a
range of Hubble type and star formation properties had, until
recently, not been performed.

Magnetic fields are expected to play an important role in several
aspects of star formation and galaxy evolution. First, magnetic fields
are a crucial consideration in the energy balance of the ISM
\citep[e.g.,][]{beck_2007}, and in particular are likely important in
determining the conditions for gravitational instability that lead to
the inital stages of star formation
\citep{mckee_ostriker_2007}. Magnetic fields are expected to be
important agents in helping to shape galactic evolution on large
scales \citep{boulares_cox_1990}, while the longevity of familiar
morphological features such as spiral arm ``spurs'' may be dependent
on the presence of ordered magnetic fields
\citep{shetty_ostriker_2006}. Finally, magnetic fields may be an
important piece of the puzzle in understanding how the disk-halo
interaction proceeds \citep[e.g.,][]{tuellmann_etal_2000}, and thus in
determining how matter and energy are redistributed throughout a
galactic disk and indeed within galaxy groups and clusters by feedback
processes.

Clearly, magnetic fields should be a major consideration in the study
of star formation and galaxy evolution. However, observational
measurements of the magnetic fields in nearby galaxies are relatively
few. A review of the available observations is not warranted here, but
some of the most recent studies, obtained with widely varying
observational setups, include those of: NGC 6946 \citep{beck_2007};
NGC 5194 \citep{berkhuijsen_etal_1997}; NGC 4254 \citep{chyzy_2007};
and the Large Magellanic Cloud \citep{gaensler_etal_2005}. Taken
together, all of the observations indicate a tendency for magnetic
fields to be oriented in spiral patterns in disk galaxies, even in
galaxies with no spiral structure visible in the gaseous or stellar
morphology \citep[e.g. NGC 4414;][]{soida_etal_2002}.
Where spiral arms are visible, the fields tend to be more
ordered in the interarm regions. Halos seem to have large-scale
magnetic fields; the ordered fields typically lie parallel to the disk in
edge-on galaxies, and then turn to a more perpendicular orientation as
distance from the midplane increases.

Information about the magnetic fields in galaxies is most efficiently
obtained using two complementary techniques. The nonthermal
synchtrotron emission generated by relativistic electrons spiraling in
a magnetic field oriented perpendicular to the line of sight (LOS) is
linearly polarized. The electric field vector of the polarized
radiation is oriented perpendicular to the magnetic field that
accelerates the source electrons, and the radiation itself is beamed
parallel to the trajectory of the ultra-relativistic electron. Thus,
the plane of polarization of the observed synchrotron radiation is
directly related to the component of the magnetic field perpendicular
to the LOS ($B_{\perp}$) in the observed object. Moreover, the
synchrotron emissivity is proportional to the product of
$B_{\perp}^{1+\alpha}$ and the relativistic electron density $n_{CR}$,
where $\alpha$ is the spectral index \citep[e.g.,][]{longair_1994}.
This makes the observed
polarized intensity itself a good tracer of $B_{\perp}$. This
straightforward correspondence is complemented by the second technique
for measuring magnetic fields: Faraday rotation. This effect is
produced when polarized radiation passes through a magnetized plasma,
which is birefringent \citep{gardner_whiteoak_1963}. The intrinsic
linear polarization angles of the radiation are rotated by a different
angle depending on the wavelength of the radiation. The effect is
characterized by the Faraday ``rotation measure''. The value of the
rotation measure ($RM$) is dependent on the electron density $n_e$ in
the magnetized plasma, and the component of the magnetic field along
the LOS ($B_{\parallel}$). The sign of RM is determined by whether
$B_{\parallel}$ points toward or away from the observer.
See \S\,\ref{subsection:rmsynthesis} for an in-depth discussion.

The Spitzer Infrared Nearby Galaxies Survey
\citep[SINGS;][]{kennicutt_etal_2003} was conceived as a
multi-wavelength Legacy program intended to address the question of
how stars form in a wide range of galactic ISM environments. The
strength of such a concerted survey campaign is that it draws together
data over the widest possible range of observing bands to provide as
much information as possible about the physical conditions in the
galaxy ISM being investigated. Gaps in the coverage are generally
covered by supplementary surveys such as The \textsc{H$\,$i} Nearby
Galaxy Survey \citep[THINGS;][]{walter_etal_2008}.

One such supplementary survey is the Westerbork SINGS survey
\citep[WSRT-SINGS;][]{braun_etal_2007}, which provides 18- and 22-cm
radio continuum data, in all four Stokes parameters, for a subset of
the SINGS galaxies (the survey selection criteria are discussed
below). Together with the SINGS survey itself, the data provided by
the WSRT supplement enable, for example, investigation into the origin
of the FIR-radio correlation \citep{murphy_etal_2006}. In this paper
and a companion work (Braun, Heald \& Beck 2009; hereafter Paper III),
we utilize the linear polarization products of the WSRT-SINGS data to
investigate the magnetic field content of the ISM in the subsample
galaxies.

Of the galaxies that make up the SINGS sample, not all are observable
with the WSRT. Because the individual antennas are arranged in a
linear east-west array, the synthesized beam is significantly extended
in the north-south direction when observing objects at low
declination. At the frequencies observed in this survey, the
synthesized beam would be $\gtrsim\,1\arcmin$ for sources at
$\delta\,<\,12.5\degr$; we therefore exclude galaxies below this
declination limit. Furthermore, in order to ensure that the galaxies
themselves are large enough on the sky that they are spatially
resolved, the additional criterion was adopted that the optical B band
diameter at a surface brightness of 25~mag~arcsec$^{-2}$,
$D_{25}\,>\,5\arcmin$. With the addition of four galaxies in the
Starburst sample of G. Rieke, a total of 34 galaxies were observed in
the WSRT-SINGS program. Twenty-eight of those galaxies are studied
here; their properties are summarized in Table \ref{table:summary}.
The columns are (1) Galaxy ID; (2) RC3 Hubble type; (3) $D_{25}$;
(4) Inclination; (5) Spiral pitch angle (from \citet{kennicutt_1981});
(6) Spiral sense (+1 for counter-clockwise, $-1$ for
clockwise); (7) Kinematic PA (measured east of north) of the receding
major axis; (8) Reference for inclination and PA values;
(9) Synthesized beam ellipticity ($a/b$, where the minor axis of the beam
is in all cases $b=15\arcsec$, and the beam position angle is $0\degr$;
(10) Noise levels in $P$; (11) Integrated flux in $P$ with an estimated
error; (12) Estimated foreground RM that applies to the target field;
(13) Integrated 1365 MHz flux in $I$ (from \citet{braun_etal_2007}).

\begin{table*}
\begin{minipage}[t]{\hsize}
\caption{Summary of Survey Galaxies.}
\label{table:summary}
\centering
\renewcommand{\footnoterule}{}
\begin{tabular}{l l c c c c c c c c c c c }
\hline\hline
Galaxy & Hubble & $D_{25}$ & Incl. & $\psi^\prime_{xy}\pm$~err & Spiral & PA & Ref\footnote{References. (1) \citet{deblok_etal_2008}; (2) \citet{paturel_etal_2003};
  (3) \citet{braun_etal_2007}; (4) \citet{bureau_carignan_2002};
  (5) \citet{kamphuis_briggs_1992}; (6) \citet{jarrett_etal_2003};
  (7) \citet{braun_1995}; (8) \citet{rc3};
  (9) \citet{tamburro_etal_2008}} & Beam &  
$\sigma_P$ & $P~\pm$~err & $RM_{FG}~\pm$~err &$I$\\
ID & Type & [arcmin] & [deg.] & [deg.] & sign & [deg.] & \ & ellipse & [$\mu$Jy/bm] & [mJy] & [rad~m$^{-2}$] & [mJy]\\
\hline
Holmberg II & Im & 7.9 & 56 &\ldots &-- & 168 & 4 & 1.06 & 11.8 & $<0.5$ & $-10~\pm~2$ & 5.5 \\ 
IC 2574 &     SABm & 13.2 & 53 &-- &+1 & 56 & 1 & 1.08 &  12.5 & $<0.6$ & $-18~\pm~4$ & 14. \\  
NGC 628 &     SAc & 10.5 & 7 & 15$\pm$2 &+1 & 25 & 5 & 3.68 &  13.5 & 23~$\pm$~2 & $-34~\pm~2$ & 200. \\  
NGC 925 &  SABd & 10.5 & 66 &25$\pm$2 & +1& 287 & 1 & 1.81 &  13.0 & $<0.6$ & $-10~\pm~2$ & 90. \\ 
NGC 2403 & SABcd & 21.9 & 63 & 21$\pm$4&+1 & 124 & 1 & 1.10 &  9.0 & 28~$\pm$~5 & $+11~\pm~30$ & 360 \\  
NGC 2841 & SAb & 8.1 & 74 &-- &$-1$ & 153 & 1 & 1.29 &  8.6 & 5.8~$\pm$~0.5 & $-6~\pm~3$ & 100 \\   
NGC 2903 & SABbc & 11.5 & 65 & 13$\pm$5&$-1$ & 204 & 1 & 2.73 &  12.6 & 14~$\pm$~1 & $+3~\pm~1$ & 460 \\ 
NGC 2976 & SAc & 5.9 & 65 &-- &$-1$ & 335 & 1 & 1.08 &  9.5 & 3.0~$\pm$~0.6 & $-34~\pm~2$ & 68. \\ 
NGC 3184 & SABcd & 7.4 & 16 & 17$\pm$3&+1 & 179 & 9 & 1.51 & 9.7 & 5~$\pm$~1 & $+19~\pm~2$ & 80. \\ 
NGC 3198 & SBc & 8.5 & 72 &-- &$-1$ & 215 & 1 & 1.40 & 9.4 &  $<0.5$ & $+7~\pm~2$ & 49. \\ 
NGC 3627 & SABb & 9.1 & 62 &-- &$-1$ & 173 & 1 & 4.45 &  15.2 & 12.5~$\pm$~1 & $+13~\pm~1$ & 500 \\  
NGC 3938 & SAc & 5.4 & 14 &12$\pm$3 &+1 & 204 & 2,3 & 1.44 &  9.1 & 6.2~$\pm$~0.5 & $+1~\pm~2$ & 80. \\ 
NGC 4125 & E6p & 5.8 & -- &-- &-- & 83\footnote{Morphologically-determined position angle} & 6 & 1.10 &  6.7 & $<0.1$ & $+19~\pm~1$ & 1.9\\  
NGC 4236 & SBdm & 21.9 & 73 &-- &$-1$ & 161 & 7 & 1.07 &  9.1 & $<0.3$ & $+15~\pm~2$ & 26. \\ 
NGC 4254 & SAc & 5.4 & 32 &22$\pm$4 &+1 & 65 & 2,3 & 4.02 &  14.5 & 24~$\pm$~2 & $-13~\pm~3$ & 510 \\
NGC 4321 & SABbc & 7.4 & 30 &15$\pm$3 &$-1$ & 159 & 2,3 & 3.67 &  13.1 & 16~$\pm$~1 & $-17~\pm~1$ & 310 \\
NGC 4450 & SAab & 5.2 & 48 &10$\pm$2 &+1 & 352 & 2,3 & 3.40 & 13.5 & $<0.2$ & $-8~\pm~1$ & 13. \\ 
NGC 4559 & SABcd & 10.7 & 65 &-- &$-1$ & 328 & 2,3 & 2.13 &  12.9 & 1.0~$\pm$~0.2 & $-5~\pm~2$ & 110\\  
NGC 4569 & SABab & 9.5 & 66 &-- &$-1$ & 23 & 2,3 & 4.39 &  13.3 & 12~$\pm$~1 & $+18~\pm~2$ & 170\\    
NGC 4631 & SBd & 15.5 & 85 &-- &-- & 86 & 2,3,8 & 1.86 &  12.6 & 40~$\pm$~2 & $-4~\pm~3$ & 1290\\ 
NGC 4725 & SABab & 10.7 & 54 &7$\pm$1 &+1 & 36 & 2,3 & 2.32 &  12.4 & 3.6~$\pm$~0.5 & $-4~\pm~4$ & 100 \\   
NGC 4736 & SAab & 11.2 & 41 &-- &+1 & 296 & 1 & 1.52 & 17.3 & 11.5~$\pm$~1 & $+1~\pm~1$ & 320 \\  
NGC 4826 & SAab & 10.0 & 65 &7$\pm$2 &+1 & 121 & 1 & 2.71 &  13.5 & 0.7~$\pm$~0.1 & $-8~\pm~2$ & 110 \\  
NGC 5033 & SAc & 10.7 & 66 &19$\pm$3 &$-1$ & 352 & 2,3 & 1.68 & 10.5 & 3.5~$\pm$~0.5 & $+9~\pm~2$ & 240 \\  
NGC 5055 & SAbc & 12.6 & 59 &11$\pm$3 &+1 & 102 & 1 & 1.49 &  10.6 & 17.5~$\pm$~1 & $-8~\pm~3$ & 450 \\ 
NGC 5194 & SABbc & 11.2 & 42 &15$\pm$2 &$-1$ & 172 & 9 & 1.36 &  9.5 & 81~$\pm$~5 & $+12~\pm~2$ & 1420 \\ 
NGC 6946 & SABcd & 11.5 & 33 &28$\pm$4 &+1 & 243 & 1 & 1.15 &  10.6 & 150~$\pm$~10 & $+23~\pm~2$ & 1700\\ 
NGC 7331 & SAb & 10.5 & 76 &14$\pm$3 &+1 & 168 & 1 & 1.77 &  11.4 & 14.5~$\pm$~1 & $-177~\pm~7$ & 590 \\ 
\hline
\end{tabular}
\end{minipage}
\end{table*}

This paper is organized as follows. We describe the observations and
data reduction steps in \S\,\ref{section:reductions}, with a
particular emphasis on describing the RM-Synthesis method
(\S\,\ref{subsection:rmsynthesis}), which is a critical component of
the analysis utilized in this work. An overview of the polarized
emission detected in each of the survey galaxies is presented in
\S\,\ref{section:overview}. For those galaxies with detected polarized
emission, a discussion of some derivable characteristics is given in
\S\,\ref{section:discussion}. Properties of the global magnetic field
geometries revealed by these observations are treated in detail in
Paper III. A more detailed study of individual
galaxies will form the basis of forthcoming work. We conclude the
paper in \S\,\ref{section:conclusions} and provide an outlook for future
investigations.

\section{Observations and Data Reduction\label{section:reductions}}

\subsection{Data collection and `standard' data reduction\label{subsection:standardreductions}}

The observational parameters of the WSRT-SINGS survey were presented
in detail by \citet{braun_etal_2007}, and we list the most relevant
points here. Each galaxy was observed for at least 12 hr in two bands
covering the ranges 1300--1432 and 1631--1763 MHz (22- and 18-cm,
respectively). The observing band was switched every 5 minutes during
an individual synthesis. In each band, the correlator was set up to
provide 512 channels separated by 312.5 kHz. Eight 20-MHz
subbands (64 channels each) were used at each observing frequency,
and the central
subband frequencies were arranged to be separated by 16 MHz. This setup
allows us to disregard frequency channels suffering from bandpass rolloff
(which affects each of the individual subbands),
and maximizes the continuity of the frequency coverage while still
providing a large total bandwidth. Data were obtained in all four
Stokes parameters.

The basic data reduction steps of each 20 MHz subband are also
discussed by \citet{braun_etal_2007}; we repeat the most relevant
details here. After careful editing of
incidental radio frequency interference (RFI) the bandpass
calibration in amplitude and phase was determined using the
calibration sources 3C147, 3C286, CTD93 and 3C138 within the {\tt AIPS}
package \citep{greisen_2003}. Relative broadband gains in the two
perpendicular linear polarizations (X and Y) were then
determined, after modifications to several key tasks ({\tt SETJY} and {\tt CALIB})
to enable the representation of source models, and the calculation of gain solutions,
with arbitrary values of the Stokes parameters $(I,Q,U,V)$. This was necessary to permit an
equivalent representation of the measured linear polarization products
(with an unchanging parallactic angle) within a software package that
normally assumes right- and left-handed circular polarization
products.  Basic polarization calibration was then accomplished by
determining the cross-polarization leakage from 3C147, under the
assumption that this source is intrinsically unpolarized. The phase
offset of the X and Y polarizations (which is assumed to
remain constant during each 12 hr track) was then determined using the
linearly polarized emission properties of either 3C286
[e.g. $(I,Q,U,V)$=(14.65,0.56,1.26,0.00) Jy near 1400 MHz] or 3C138. In
cases where both 3C286 and 3C138 were observed bracketing the 12~hr
target track, it was possible to determine the consistency of the phase
offset, which was found to be constant to better than 1--2
degrees. After a final check that the correct Stokes parameters were
recovered for all calibration sources (both polarized and unpolarized),
the calibrated data were exported from the {\tt AIPS} package. Further
refinement of the
polarization calibration was accomplished via self-calibration of each
20 MHz subband within the Miriad package \citep{sault_etal_1995}
using the detected emission in each target field in Stokes $I, Q$ and
$U$. This step corrects for time-variable instrumental or ionospheric
phase errors.

Following these reduction steps, the $Q$ and $U$ maps in each narrowband
frequency channel (of 312.5 kHz) were imaged individually. They were
{\tt CLEAN}ed within mask regions derived from smoothed versions of
the total Stokes $I$ map in each subband. The same restoring beam size was 
used at all channel frequencies; see Table
\ref{table:summary} for the effective resolution in each galaxy. Primary
beam corrections were applied to each channel map individually. The primary
beam correction was performed using the standard WSRT model [the
primary beam response is approximated with the function
$\cos^6(c{\nu}r)$, where $c=68$ at L-band frequencies, $\nu$ is the
frequency in GHz, and $r$ is the distance from the pointing center in
radians]. An improved primary beam correction for the WSRT has been
developed by \citet{popping_braun_2008} but is little different from
the standard treatment at the small field angles which are considered
here. After primary beam correction, the channel maps were arranged
into cubes, and analyzed with the RM-Synthesis technique (described in
\S\,\ref{subsection:rmsynthesis}). Individual channel maps are used
as input to the RM-Synthesis software because the technique can be thought of
as a technique for determining the Faraday rotation measure which maximizes the
polarized signal-to-noise ratio across the full frequency band.

Stokes $V$ images were generated, and the intensity histograms in those
images are Gaussian, with an rms of about $20\,\mu\mathrm{Jy\,beam}^{-1}$.
In some fields, very bright continuum sources far from the field center have
instrumental circular polarization, at the $V/I\lesssim0.5\%$ level.

\subsection{RM Synthesis\label{subsection:rmsynthesis}}

As discussed above, the effect of Faraday rotation is to change the
intrinsic polarization angle of the radiation ($\chi_0$) by an amount
depending on the wavelength of the radiation. More specifically, the
observed polarization angles after Faraday rotation are
\begin{equation}
\chi\,=\,\chi_0+\phi\,\lambda^2,
\label{equation:RM0}
\end{equation}
where $\lambda$ is the wavelength, and the rotation measure $RM$ has
been replaced by a more general quantity $\phi$, the ``Faraday
depth.'' The value of $\phi$ is related to the properties of the
Faraday rotating plasma by the equation
\begin{equation}
\phi\propto\int_{\mathrm{source}}^{\mathrm{telescope}}n_e\,\vec{B}\cdot\,d\vec{l},
\label{equation:RM1}
\end{equation}
where $\vec{B}$ is the magnetic field, $\vec{l}$ is the distance along
the LOS, and $n_e$ is the electron density. When $\vec{B}$ is
expressed in $\mu$Gauss, $\vec{l}$ in pc, and $n_e$ in cm$^{-3}$, the
proportionality constant is 0.81, and the units of $\phi$ are
$\rpms{}$. From equation \ref{equation:RM1}, it can be seen that the
Faraday depth expresses the depth of Faraday rotating plasma between
the source and the telescope. The dot product indicates that it is
only the component of the magnetic field along the LOS
($B_{\parallel}$) that contributes to the integral. The sign of $\phi$
is taken to be positive for a magnetic field pointing toward the
observer. The magnitude of $\phi$ is traditionally determined by
fitting a linear relationship between $\chi$ and $\lambda^2$
\citep[e.g.,][]{ruzmaikin_sokoloff_1979}. However, this approach has
well-documented problems such as $n\pi$ ambiguities. Sophisticated
techniques have been developed to overcome these difficulties
\citep[e.g. Pacerman,][]{dolag_etal_2005,vogt_etal_2005}, but these
still suffer from the same fundamental problem. Moreover, by fitting
the change in the observed polarization angle with $\lambda^2$, one is
faced with the implicit constraint that the polarized signal must be
bright enough at each observing frequency to allow a significant
fit. For the faintest polarized emission this would not be possible. A
different method for determining the effect on the polarized radiation
produced by the magnetized plasmas along the LOS, first described by
\citet{burn_1966} and now called the RM-Synthesis method
\citep{brentjens_debruyn_2005}, can overcome the weaknesses in the
traditional techniques for determining RM, and will be used in this
paper.

Equation \ref{equation:RM0} is valid only in physical situations where
all of the polarized emission is observed at a single Faraday depth
$\phi$. In more complicated circumstances
\citep[for instance, emission that
arises both beyond and between two distinct Faraday rotating clouds
along the LOS; for an in-depth discussion see][]{sokoloff_etal_1998},
the simple relation is no longer valid. By expressing
the polarization vector as an exponential ($P=p\,e^{2i\chi}$), using
equation \ref{equation:RM0} for $\chi$, and integrating over all
Faraday depths, \citet{burn_1966} shows that
\begin{equation}
P(\lambda^2)=\int_{-\infty}^{+\infty}F(\phi)\,e^{2i\phi\lambda^2}\,d\phi,
\label{equation:phi}
\end{equation}
where $P(\lambda^2)$ is the (complex) observed polarization vector
[$P(\lambda^2)=Q(\lambda^2)+iU(\lambda^2)$], and $F(\phi)$, the
``Faraday dispersion function,'' describes the {\em intrinsic}
polarization vector at each Faraday depth.

\begin{figure}
\resizebox{\hsize}{!}{\includegraphics{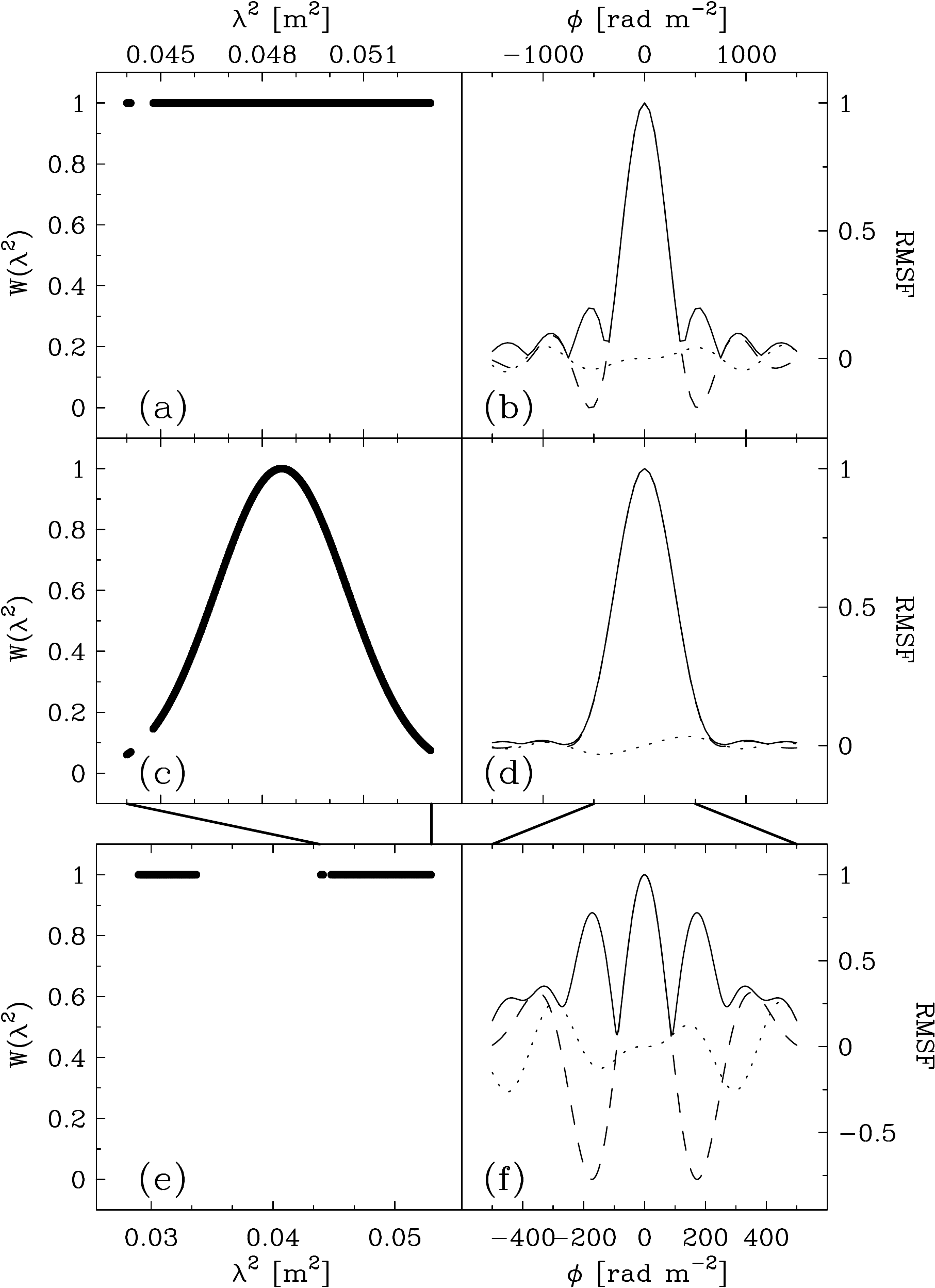}}
\caption{RMSFs corresponding to the frequency coverage in the
  observations of NGC 628. RMSF values were calculated using the 22cm
  band alone (untapered: \emph{a,b}; tapered: \emph{c,d}) and in the
  combination of the 18cm and 22cm bands (\emph{e,f}). In each row,
  the left panel shows the value of the weight function $W(\lambda^2)$
  for each of the sampled frequency channels, and the right panel
  shows the corresponding RMSF (real part: \emph{long dashed lines};
  imaginary part: \emph{dotted lines}; absolute value: \emph{solid
    lines}). Matching ranges of $\lambda^2$ are illustrated by
  diagonal lines between panels \emph{c} and \emph{e}, and matching
  ranges of $\phi$ between panels \emph{d} and \emph{f}.}
\label{figure:RMSFs}
\end{figure}

Under the assumption that $\chi_0$ is constant for all $\phi$, the
Fourier transform-like eqn. \ref{equation:phi} can be inverted to give
an expression for the Faraday dispersion function:
\begin{equation}
F(\phi)=\int_{-\infty}^{+\infty}P(\lambda^2)\,e^{-2i\phi\lambda^2}\,d\lambda^2.
\label{equation:phiinv}
\end{equation}
Everything on the
right-hand side of eqn. \ref{equation:phiinv} is observable. However,
we only measure discrete ({\em positive}) values of $\lambda^2$. The
form that is used in practice \citep[e.g.,][]{brentjens_debruyn_2005}
is therefore expressed as a discrete sum,
\begin{equation}
F(\phi)\,\approx\,K\,\sum^N_{i=1}\,W_i\,P_i\,e^{-2i\phi(\lambda_i^2-\lambda_0^2)},
\label{equation:phiinvsum}
\end{equation}
where $W_i$ are weights which are allowed to differ from unity,
and the normalization factor $K$ is
the inverse of the discrete sum over $W_i$. Note that the term
$\lambda_0^2$ has been added to the
exponential. \citet{brentjens_debruyn_2005} demonstrate that when
$\lambda_0^2$ is taken to be the weighted mean of the $N$ individual
observed $\lambda_i^2$, a better behaved response function
results. That response function, or rotation measure spread function
(RMSF)\footnote{This function was originally referred to as the
  rotation measure transfer function (RMTF) by
  \citet{brentjens_debruyn_2005}, but has since been relabeled to more
  accurately reflect its mathematical relationship to the Faraday
  dispersion function.}, is formally given by
\begin{equation}
R(\phi)\,\approx\,K\,\sum_{i=1}^N\,W_i\,e^{-2i\phi(\lambda_i^2-\lambda_0^2)}
\label{equation:RMSF}
\end{equation}
and is conceptually equivalent to the dirty beam encountered when
performing image synthesis with an array of radio telescopes. Just as
a radio interferometer discretely samples $uv$-space, here we
discretely sample $\lambda^2$-space. Examples of the RMSF, specific to
the observations presented in this paper, are shown in
Fig. \ref{figure:RMSFs}.

Previous rotation measure experiments have had to rely on relatively few
measurements in frequency space. But with modern correlator backends
like the one at the WSRT, the technique of determining $F(\phi)$ as
shown in equation \ref{equation:phiinvsum} is made possible. The
practical aspects of this technique have been developed by
\citet{brentjens_debruyn_2005}. We use software developed by
M. Brentjens to perform the inversion shown in
Eqn. \ref{equation:phiinvsum} and obtain a reconstruction of
$F(\phi)$. The software takes cubes of Stokes $Q$ and $U$ images
in single frequency channels as input, along with a specification of the
frequency at each plane of the cubes. As
output, cubes of Stokes $Q$ and $U$ in planes of constant $\phi$ are
obtained. Simply put, the inversion amounts to the computation of the
implied values of $Q$ and $U$ for a whole series of trial values of the
Faraday depth, $\phi$. In this way, the coherent sensitivity of the
entire observing band to polarized emission is retained, irrespective
of possible Faraday rotation within the band, as long as such rotation
is well resolved by the $\lambda^2$ sampling.

The polarization vectors described by the values of $Q$ and $U$ in each
plane can be thought of as having been corrected for Faraday rotation
-- but note that the vectors have been derotated to a common non-zero
value of $\lambda^2$, namely to $\lambda^2\,=\,\lambda_0^2$, as shown
in Eqn. \ref{equation:phiinvsum}. Hence, to obtain the {\em intrinsic}
polarization angle at each value of $\phi$, multiplication of our
reconstructed $F^{\prime}(\phi)=Q(\phi)+iU(\phi)$ by
$e^{-2i\phi\lambda_0^2}$ must be performed. Further explanation
regarding this detail is provided in Appendix \ref{appendix:rmclean}.

The frequency sampling provided by the WSRT-SINGS survey gives
sensitivity to polarized emission up to a maximum Faraday depth of
$|\phi_{\mathrm{max}}|\,\approx\,\sqrt{3}/\delta\lambda^2\,\approx\,\rpms{1.7\times10^5}$
\citep[see][]{brentjens_debruyn_2005}, where $\delta\lambda^2$ refers
to the channel separation. A search for large rotation measure
emission was performed for each of the galaxy fields, by performing
RM-Synthesis on the observed $Q$ and $U$ cubes in the range
$|\phi|\,<\,\rpms{1.7\times10^5}$ (albeit with coarse $\phi$
sampling). No emission at high values of Faraday depth was found in
any of the target fields. Next, RM-Synthesis was performed on the 22cm
data alone, from $\rpms{-1500}$ to $\rpms{+1500}$ with fine sampling
($\rpms{50}$). Given the $\lambda^2$ width of the 22cm band, the
$\phi$ resolution element (FWHM) is $\rpms{450}$. The first sidelobe
of the RMSF is at 24\% of the main lobe. The sidelobe level can be
reduced, at the expense of lower $\phi$ resolution, by tapering in
$\lambda^2$ space (by allowing $W_i$ to deviate from unity in
equation \ref{equation:phiinvsum}, and as illustrated in Figure
\ref{figure:RMSFs}). After tapering with a Gaussian with
$\sigma\,=\,\frac{1}{2}(\lambda_{\mathrm{max}}^2-\lambda_{\mathrm{min}}^2)$,
the width of the main lobe is $\rpms{650}$, and the first sidelobe is
reduced to 2\% of the RMSF peak. Increasing the denominator in the
tapering function serves to further decrease the sidelobe level, while
increasing the width of the main lobe. After testing a series of different
tapers, this particular choice was selected as a reasonable tradeoff
between sidelobe height and RMSF width. The cubes produced in this way can be
used as a (very) low resolution verification of complicated $F(\phi)$
spectra.

RM-Synthesis was also performed on the combination of the 18cm and
22cm data. The results of this operation were used for most of the
subsequent analysis. Together, the two bands provide a $\phi$
resolution of $\rpms{144}$. This is comparable to the maximum Faraday
depth to which about 50\% sensitivity to the polarized intensity is
retained of about $\pi/\lambda^2_{min}~=~\rpms{110}$. However, due to the
large gap in frequency coverage between the two bands, the RMSF has
large sidelobes, as shown in Figure \ref{figure:RMSFs}. The first
sidelobes are at about the 78\% level in P, which can potentially
cause serious confusion, particularly in cases where polarized
emission is detected at multiple $\phi$ along a single LOS. This
difficulty can be alleviated by using a deconvolution technique
similar to the H\"ogbom {\tt CLEAN} algorithm
(\S\,\ref{subsection:rmclean}).

\subsection{Faraday dispersion function deconvolution\label{subsection:rmclean}}

Once the RM-Synthesis was performed for each field, the $F(\phi)$
spectra were deconvolved using a variation of the H\"ogbom {\tt
  CLEAN}, as outlined by \citet{brentjens_2007}. The deconvolution is
complex-valued and operates along the $\phi$ dimension, which is the
third axis of the $Q(\phi)$ and $U(\phi)$ cubes produced by the
RM-Synthesis technique. The steps of the procedure, called {\tt
  RM-CLEAN}, are described in detail in Appendix
\ref{appendix:rmclean}. Briefly, one iteratively subtracts scaled
versions of the RMSF from the reconstructed Faraday dispersion function
until the noise floor is reached, after which a smoothed
representation of the ``{\tt CLEAN} model'' is used as the approximate
true Faraday dispersion function. In this paper, we take the {\tt
  RM-CLEAN} cutoff to be equal to the noise in the individual
$Q(\phi)$ and $U(\phi)$ maps, and the gain factor is 0.1 (see Appendix
\ref{appendix:rmclean} for a more extensive description of these parameters).

\begin{figure}
\resizebox{\hsize}{!}{\includegraphics{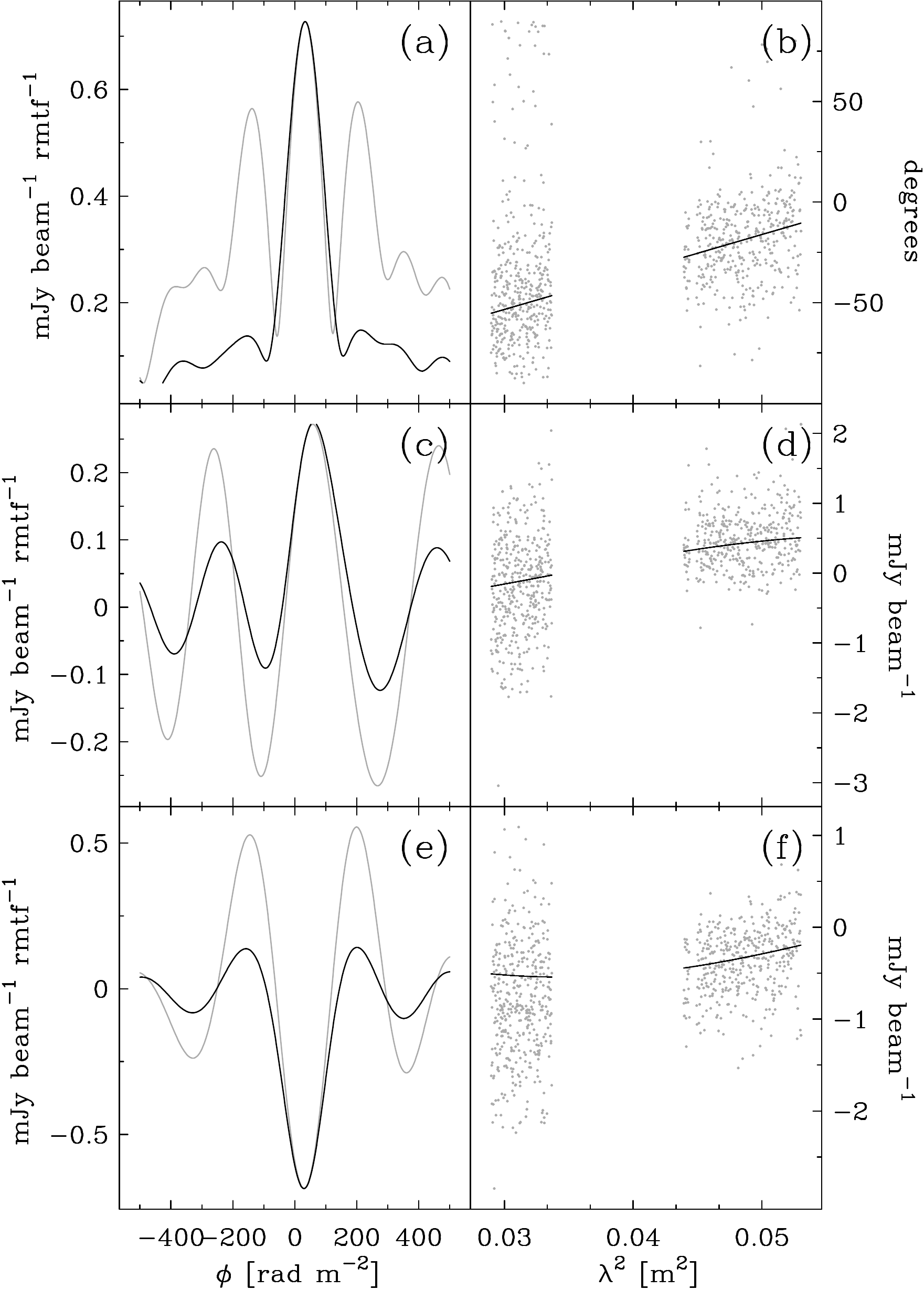}}
\caption{Demonstration of the {\tt RM-CLEAN} process for a relatively
  bright point source in the field of NGC 4125. Top: $P$, middle: $Q$,
  bottom: $U$. Gray lines are the dirty spectra; black lines are the
  cleaned spectra. A restoring RMSF with FWHM $\rpms{144}$ was used to
  produce the deconvolved spectra. The resulting $F(\phi)$ spectrum
  shows a single component at a central $\phi=\rpms{+33}$.}
\label{figure:cleanex}
\end{figure}

\begin{figure}
\resizebox{\hsize}{!}{\includegraphics{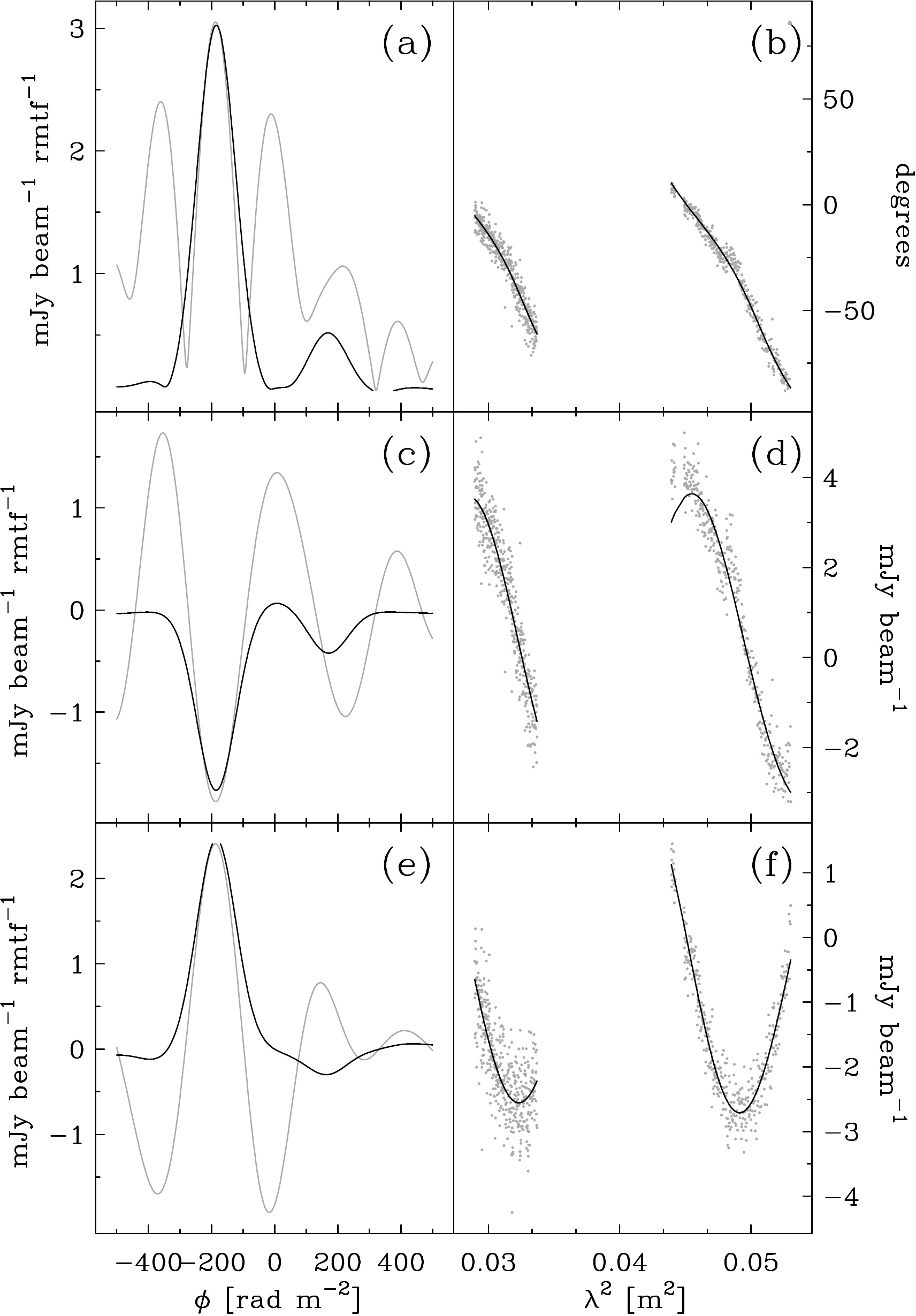}}
\caption{Demonstration of the {\tt RM-CLEAN} process for a relatively
  bright point source in the field of NGC 7331. Top: $P$, middle: $Q$,
  bottom: $U$. Gray lines are the dirty spectra; black lines are the
  cleaned spectra. A restoring RMSF with FWHM $\rpms{144}$ was used to
  produce the deconvolved spectra. The resulting $F(\phi)$ spectrum
  shows two components at central $\phi=\rpms{-185,+170}$.}
\label{figure:cleanex2}
\end{figure}

Examples of the result of running {\tt RM-CLEAN} on dirty $F(\phi)$
spectra are shown in Figures \ref{figure:cleanex} and
\ref{figure:cleanex2}. In Fig. \ref{figure:cleanex}, a single-valued
$F(\phi)$ spectrum has been {\tt RM-CLEAN}ed. The only benefit is that
the sidelobe structure has been significantly reduced. Note that
the deconvolution routine is unable to improve the $\phi$ resolution,
which is determined by the spread of sampled frequencies. In the
right-hand panels, the data are compared to the $\lambda^2$
representation of the {\tt RM-CLEAN} components found during the
procedure. A solution equivalent to the best linear fit to
$\chi$-vs-$\lambda^2$ has been determined. That this type of
deconvolution is mathematically identical to a least-squares fit in
the inverse Fourier domain has been shown by \citet{schwarz_1978}. In
Fig. \ref{figure:cleanex2}, a more complicated $F(\phi)$ is
shown. Note that in cases such as this, where multiple structures are
detected, the location of the peak of the fainter component is shifted
relative to its true position because of confusion with sidelobes from
the brighter component. Particularly in such cases, deconvolution is
required to recover source parameters.

Final on-axis sensitivities in the deconvolved $F(\phi)$
cubes are listed in Table~\ref{table:summary} for each galaxy. In view
of the combined frequency coverage contributing to $P$ (1300--1432 and
1631--1763 MHz), the effective center frequency is about 1530~MHz.

\subsection{Analysis of deconvolved Faraday depth cubes\label{subsection:analysis}}

The spatial distribution of polarized emission was determined by
selecting the peak in each
$P(\phi)\,\equiv\,||F(\phi)||$ spectrum. The grayscale maps of
these images are shown in Figure \ref{figure:images:a}, with overlaid
contours of the 22cm Stokes-I maps for comparison. Since polarized
intensity has Ricean, rather than Gaussian statistics
\citep[see][]{wardle_kronberg_1974}, some care needs to be taken in
determining the integrated value of $P$. In the absence of source
signal, the noise has a Rayleigh distribution with a mean (the
``Ricean bias'') and variance of
\begin{equation}
{\langle}P{\rangle}_n\,=\,\sigma_{Q,U}\sqrt{\pi \over 2}\,=\,1.25\cdot\sigma_{Q,U},
\end{equation}
and
\begin{equation}
\sigma_P\,=\,\sqrt{4-\pi \over 2}\,\sigma_{Q,U}\,=\,0.66\cdot\sigma_{Q,U},
\end{equation}
for an RMS noise in $Q$ and $U$ of $\sigma_{Q,U}$. In the case of high
signal-to-noise, the noise statistics become Gaussian with the mean
$P$ in agreement with its noise-free value and
$\sigma_P\,=\,\sigma_{Q,U}$.

We have carefully determined the noise level and Ricean bias in
each target. To do this,
we have determined the mean and variance
in each image of peak $P$ within a central, polygonal region (free of
emission). The variances that we measure (and list in
Table \ref{table:summary}) are in good agreement with the expectation
noted above, $\sigma_P\,=\,0.66\cdot\sigma_{Q,U}$, for a Rayleigh
distribution. However, the bias values that we measure are
in all cases enhanced by a factor of about two over the simple expectation,
${\langle}P{\rangle}_n\,=\,1.25\cdot\sigma_{Q,U}$. The cause of this
high Ricean bias level is that the peak value of $P(\phi)$
has been extracted from a cube covering Faraday depths
between $-500$ and $\rpms{+500}$ with an effective resolution of about
$\rpms{144}$, and has thus been chosen from some 7 independent
samples. The Rayleigh distribution function is given by
\begin{equation}
D(r)\,=\,1\,-\,\exp\left({-r^2 \over 2\sigma_{Q,U}^2}\right),
\end{equation}
where $r$ represents the flux in a given sample. In a Faraday dispersion
function which contains only noise (no signal), the expectation
value for the largest (${\hat r}$) of N independent samples occurs for
\begin{equation}
\exp\left({-{\hat r}^2 \over 2\sigma_{Q,U}^2}\right)\,=\,{1 \over N},
\end{equation}
or 
\begin{equation}
{\hat r}\,=\,\sigma_{Q,U}[2\,\ln(N)]^{0.5}.
\end{equation}
This should be a good estimator of the Ricean bias in our maps. Since
we have $N=7$, we obtain ${\hat r}\,=\,2.0\cdot\sigma_{Q,U}$, in good
agreement with what is measured for the on-axis background level in all
cases except NGC 6946 and NGC 7331, where patchy foreground emission
from the Galaxy is apparent in the fields. The mean noise floor
increases radially away from the field center due to the primary beam
correction which has been applied to each original frequency channel
during processing. For the purposes of display and the measurement of
azimuthal trends, we have subtracted out a primary beam-corrected
noise floor from the peak $P$ images shown in Figure \ref{figure:images:a}.

For determination of the integrated $P$ (or useful limits on $P$) for
each target (as listed in Table~\ref{table:summary}), we have not
carried out a spatial background subtraction of the noise floor, but
instead have simply blanked the images at a level of
3$\sigma_{Q,U}\,\sim\,4.5\sigma_P$. At these brightnesses, the Ricean
bias has already declined to below about 5\%
\citep{wardle_kronberg_1974}, so that no further bias correction of
the integrated $P$ was applied. The $P$ emission was integrated within the
smallest possible polygonal region which enclosed the region of
significant target emission while excluding any apparent background
sources. For comparison, we also list the integrated $I$ of each
target at 1365 MHz from \citet{braun_etal_2007}.

Polarized emission maps were also produced (using only the 22cm data)
for the purpose of determining the polarized fraction in these
galaxies. The peak polarized intensity was extracted for each spatial
pixel, and then divided by the corresponding 22cm Stokes-I value. Clip
levels were set at 4 times the noise level in both maps. Thus,
polarization fraction estimates are not available for the faintest
emission detected in the sample. The polarized fraction values are 
discussed in \S\,\ref{subsection:notes}.

In order to determine the Faraday depth at the peak of the $P(\phi)$
spectra, we fit a parabola to the top three points in the oversampled,
{\tt RM-CLEAN}ed $P(\phi)$ spectra. The result is called $\hat{\phi}$.

The polarization angle at the Faraday depth thus determined was obtained via
\begin{equation}
\chi_0\,=\,\frac{1}{2}\arctan\left(\frac{U(\hat{\phi})}{Q(\hat{\phi})}\right).
\end{equation}
We refer to this as the intrinsic polarization angle ($\chi_0$)
because the effect of Faraday rotation has already been corrected for, via
the RM-Synthesis technique. The magnetic field orientation is obtained
by simply rotating the polarization angle by $90\degr$.

Errors associated with the magnetic field orientations were estimated
by propagating errors through the mathematical operations required to
calculate the quantity. Since
\begin{equation}
\chi_0\,=\,\chi\,-\,\hat{\phi}\lambda_0^2,
\end{equation}
error propagation yields the uncertainty in our determination of the
intrinsic polarization angle,
\begin{equation}
\sigma_{\chi_0}^2\,=\,\sigma_{\chi}^2\,+\,\lambda_0^4\sigma_{\hat{\phi}}^2.
\end{equation}
The quantity $\sigma_{\chi}$, the uncertainty in the observed (Faraday
rotated) polarization angle can be shown to be given by
\citep{brentjens_2007}
\begin{equation}
\sigma_{\chi}^2\,=\,\frac{1}{4}\frac{\sigma^2}{|P|^2},
\end{equation}
where $\sigma\,=\,\sigma_Q\,\approx\,\sigma_U$ is the noise in the
individual $Q(\lambda^2)$ and $U(\lambda^2)$ maps. The uncertainty in
$\hat{\phi}$ can be estimated by propagating errors in the equation
for fitting a parabola to the peak of the $P(\phi)$ profile, and
is dominated by the RMSF resolution. We have
calculated these uncertainties for all of the galaxies which have
magnetic field vectors plotted in Figure \ref{figure:images:a}, and are
typically $\lesssim\,10\degr$ in locations where the brightest $P$
emission is detected.

Note that magnetic field orientations are easily interpreted only if
the deconvolved $F(\phi)$ spectrum is single-valued. If there is more
than one $\phi$ component, then the magnetic field orientations
determined in this way only apply to the brightest component of
polarized emission. The deconvolved Faraday depth cubes
were analyzed to identify locations where multiple Faraday depths might be
present (\S\,\ref{subsection:extprof}).
These are noted throughout the paper, where appropriate.

\section{Overview\label{section:overview}}

Here, we summarize the main
features of interest observed for each galaxy field based on a comparison of
the polarized and total continuum brightness, together with the
(Faraday rotation corrected) magnetic field orientation shown in
Fig.~\ref{figure:images:a}, and the Faraday depth where peak polarized
intensity is detected in Fig.~\ref{figure:pkphi}.

\subsection{Rotation measures of discrete background sources\label{subsection:backgroundRM}}

In addition to the primary targets of our program, each observed field
also contains a number of background sources with significant
polarized brightness. While many of these sources are unresolved at
the modest angular resolution of our study ($\ge$15''), a significant
number are also resolved into the classical edge-brightened double
morphology associated with high luminosity radio galaxies
\citep[e.g.][]{miley_1980}. Lobe separations of 1--2 arcmin are
common, while a handful of objects with 5--10 arcmin angular size are
detected. We have determined the Faraday depth and the associated
error for the significantly detected polarized sources in the central
$34'\times 34'$ of our fields, making a particular note of the source
morphology and classifying sources as unresolved, double, triple
(double plus core), extended or complex. The individual lobes of
double radio sources were measured separately where practical. The
rotation measures listed in Table~\ref{table:fieldRM} were determined
for each source from a plot of Faraday depth versus polarized
brightness within a rectangular box that isolated the source
(component). The listed RM is that of the peak in $P$ while the error
corresponds to the HWHM of the distribution. The columns are
(1) Galaxy field; (2) Discrete source position; (3) Morphology;
(4) Rotation measure with estimated error. Morphologies are classified
as unresolved: UNR, double: DBL, triple: TRPL, complex: CMPLX and
extended: EXT. For DBL and TRPL sources, the individual lobes
(N, S, E or W) are measured where possible, always begining with the
brightest one (in Stokes P). The weaker lobe is prefaced by its flux
ratio with respect to the brighter. When the source is (possibly)
affected by the target galaxy disk the morphology is further flagged
as FG(?).

\begin{sidewaystable*}
\begin{center}
\caption{Discrete Source Rotation Measures.}
\label{table:fieldRM}
\begin{tabular}{l l r r l l r r l l r r}
\hline\hline
Galaxy Field& (RA, Dec)$_{\mathrm{J}2000}$ & Morphology & RM & Galaxy Field & (RA, Dec)$_{\mathrm{J}2000}$ & Morphology & RM & Galaxy Field & (RA, Dec)$_{\mathrm{J}2000}$ & Morphology & RM\\
\hline
Holmberg II & 08:16:06, 70:47:00 & UNR & $-18\pm 2$ & NGC 3627 & 11:19:28, 13:02:50 & TRPL W & $+13\pm 1$ & NGC 4725 & 12:49:17, 25:33:10 & UNR & $+15\pm 2$\\
            & 08:17:24, 70:38:03 & DBL S & $-13\pm 2$ &         &                    & 0.15 E & $+23\pm 3$ &          & 12:50:40, 25:28:05 & DBL E & $+4\pm 4$\\
            &                    & 0.95 N & $-8\pm 5$ &         & 11:20:41, 13:05:20 & UNR & $+3\pm 2$ &          & 12:50:49, 25:35:10 & UNR & $+4\pm 3$\\
            & 08:19:18, 70:55:00 & EXT & $+27\pm 5$ &         & 11:21:21, 12:51:50 & UNR & $+23\pm 3$ &          & 12:51:18, 25:31:05 & UNR & $+2\pm 2$\\
            & 08:20:16, 70:52:15 & EXT & $-8\pm 2$ & NGC 3938 & 11:52:13, 44:09:05 & DBL N & $+1\pm 2$ & NGC 4736 & 12:50:55, 41:23;00 & DBL W & $+1\pm 1$\\
            & 08:20:25, 70:53:10 & TRPL S & $-8\pm 4$ &          &                    & 0.6 S & $+7\pm 2$ &          & 12:51:17, 40:57:30 & DBL N & $-8\pm 10$\\
IC 2574 & 10:26:28, 68:19:10 & UNR & $-12\pm 3$ &          & 11:52:28, 44:21:50 & CMPLX & $+6\pm 2$ &          & 12:51:50, 41:02:10 & DBL W & $-8\pm 5$\\
        & 10:26:54, 68:24:35 & UNR & $-24\pm 3$ &          & 11:53:10, 44:14:25 & UNR (D?)& $+2\pm 2$ & NGC 4826 & 12:56:09, 21:43:34 & DBL E & $-8\pm 2$\\
        & 10:29:17, 68:13:20 & UNR & $-17\pm 2$ & NGC 4125 & 12:07:03, 65:24:35 & DBL S & $+20\pm 2$ &          &                    & 0.15 W & $-1\pm 3$\\
NGC 628 & 01:35:42, 15:37:10 & DBL W & $-36\pm 3$ &          & 12:07:33, 65:18:10 & UNR & $+31\pm 3$ &          & 12:56:19, 21:42:25 & UNR & $-13\pm 4$\\
        & 01:36:16, 15:41:40 & DBL N & $-28\pm 3$ &          & 12:07:37, 65:15:35 & UNR & $0\pm 2$ &          & 12:56:59, 21:52:35 & UNR & $+8\pm 3$\\
        & 01:36:58, 15:44:20 & UNR FG? & $-39\pm 3$ &          & 12:07:51, 65:21:20 & DBL E & $+16\pm 2$ &          & 12:57:26, 21:47:30 & UNR & $+4\pm 4$\\
        & 01:37:43, 15:42:35 & DBL E & $-38\pm 4$ &          & 12:08:02, 65:10:10 & DBL S & $+20\pm 2$ & NGC 5033 & 13:12:05, 36:23:20 & UNR & $+3\pm 1$\\
        &                    & 0.5 W & $-23\pm 3$ &          & 12:08:10, 65:00:30 & UNR & $+32\pm 2$ &          & 13:12:25, 36:40:50 & UNR & $+13\pm 1$\\
NGC 925 & 02:28:23, 33:18:45 & DBL W & $-8\pm 3$ & NGC 4236 & 12:15:30, 69:31:50 & DBL W & $+16\pm 1$ &          & 13:13:16, 36:24:50 & EXT & $+9\pm 2$\\
        & 02:28:25, 33:26:15 & UNR & $-18\pm 3$ &          &                    & 0.2 E & $+20\pm 5$ &          & 13:13:39, 36:50:20 & UNR & $+16\pm 2$\\
        & 02:28:26, 33:21:10 & UNR & $-12\pm 3$ &          & 12:16:35, 69:28:15 & DBL S FG & $-3\pm 2$ &          & 13:14:18, 36:49:15 & UNR & $-3\pm 2$\\
NGC 2403 & 07:34:18, 65:28:05 & UNR & $+13\pm 3$ &          & 12:16:36, 69:33:40 & UNR   & $+23\pm 2$ &          & 13:14:44, 36:39:00 & UNR & $-29\pm 1$\\
         & 07:38:00, 65:49:05 & UNR & $-22\pm 3$ &          & 12:18:50, 69:34:35 & DBL S & $+13\pm 2$ & NGC 5055 & 13:15:18, 41:49:25 & DBL N/S & $-5\pm 2$\\
         & 07:38:45, 65:37:45 & UNR & $+42\pm 2$ &          &                    & 0.4 N & $+17\pm 5$ &          & 13:15:33, 41:56:15 & EXT & $-18\pm 3$\\ 
NGC 2841 & 09:20:56, 51:13:50 & UNR & $-6\pm 3$ & NGC 4254 & 12:18:10, 14:15:40 & DBL E & $-13\pm 3$ &          & 13:16:25, 41:51:30 & DBL S & $-13\pm 2$\\
         & 09:21:18, 51:02:45 & UNR & $-28\pm 2$ &          &                    & 0.7 W & $0\pm 5$ &          &                    & 0.5 N & $+1\pm 4$\\
         & 09:21:25, 50:46:15 & DBL S & $-12\pm 3$ &          & 12:19:13, 14:40:10 & UNR & $-11\pm 3$ &          & 13:16:36, 42:08:55 & UNR & $-10\pm 2$\\
         & 09:22:27, 50:53:50 & DBL E & $-25\pm 2$ & NGC 4321 & 12:21:50, 15:36:10 & UNR & $-12\pm 2$ & NGC 5194 & 13:29:34, 46:58:50 & DBL E & $+14\pm 1$\\
         & 09:23:47, 51:02:40 & DBL E & $-5\pm 3$ &          & 12:23:04, 15:36:50 & UNR & $-8\pm 2$ &          & 13:29:41, 47:17:35 & UNR FG? & $+20\pm 1$\\
         &                    & 0.6 W & $-6\pm 4$ &          & 12:23:15, 15:41:40 & DBL W & $-7\pm 1$ &          & 13:30:16, 47:10:25 & EXT FG & $+28\pm 4$\\
NGC 2903 & 09:31:00, 21:35:30 & DBL W & $+20\pm 2$ &          &                    & 0.75  E & $-7\pm 2$ &          & 13:30:45, 47:03:10 & EXT & $+17\pm 2$\\
         & 09:31:13, 21:28:00 & DBL W & $+4\pm 2$ & NGC 4450 & 12:27:35, 17:07:00 & DBL E & $-8\pm 5$ &          & 13:31:25, 47:13:10 & DBL W & $+9\pm 1$\\
         &                    & 0.6 E & $+2\pm 2$ &          &                    & 0.95 W & $-10\pm 5$ &          &                    & 1.0 E & $+3\pm 1$\\
         & 09:31:37, 21:34:20 & UNR & $+8\pm 2$ &          & 12:28:15, 17:05:00 & TRPL SE & $-8\pm 1$ & NGC 6946 & 20:33:09, 60:00:30 & DBL E & $-14\pm 2$\\
         & 09:32:23, 21:36:20 & UNR & $-8\pm 2$ &          &                    &      SW & $-15\pm5$ &          &                    & 0.4 W & $+67\pm 5$\\
NGC 2976 & 09:46:18, 67:49:25 & UNR & $-44\pm 4$ &          &                    & 0.4 N & $+12\pm2$ &          & 20:35:19, 60:02:05 & DBL FG? & $+23\pm 2$\\
         & 09:46:37, 67:46:30 & UNR & $-37\pm 8$ & NGC 4559 & 12:35:48, 28:05:50 & UNR & $+4\pm 1$ &          & 20:36:09, 59:53:20 & EXT & $+14\pm 2$\\
         & 09:48:12, 67:44:15 & UNR & $-26\pm 3$ &          & 12:35:58, 27:43:15 & DBL E & $-6\pm 3$ &          & 20:36:15, 60:07:30 & UNR FG? & $+56\pm 3$\\
         & 09:48:35, 67:53:10 & DBL E & $-34\pm 2$ &          &                    & 1.0  W & $-4\pm 3$ & NGC 7331 & 22:35:51, 34:30:10 & UNR & $-143\pm 3$\\
         & 09:48:41, 68:04:15 & UNR & $+24\pm 6$ &          & 12:36:21, 27:57:30 & UNR & $+6\pm 2$ &          & 22:35:55, 34:14:25 & DBL N & $-169\pm 1$\\
NGC 3184 & 10:17:05, 41:15:10 & DBL W & $+19\pm 2$ &          & 12:37:05, 27:55:05 & UNR & $+8\pm 2$ &          &                    & 0.7 S & $-153\pm 2$\\
         & 10:17:23, 41:30:50 & UNR & $+29\pm 2$ &          & 12:37:12, 27:44:30 & UNR & $+5\pm 2$ &          & 22:35:55, 34:18:40 & UNR & $-10\pm 4$\\
         & 10:18:19, 41:09:40 & UNR & $+12\pm 3$ & NGC 4569 & 12:36:38, 12:58:55 & DBL W & $+18\pm 2$ &          & 22:36:53, 34:20:45 & DBL N & $-184\pm 2$\\
         & 10:19:01, 41:27:35 & UNR & $+11\pm 3$ &          &                    & 0.2 E & $+12\pm 2$ &          &                    & 0.2 S & $-176\pm 3$\\
         & 10:19:09, 41:41:25 & UNR & $+2\pm 2$ &          & 12:36:52, 13:01:05 & UNR & $+13\pm 2$ &          & 22:36:58, 34:16:55 & UNR & $-178\pm 3$\\
NGC 3198 & 10:19:17, 45:28:40 & UNR & $+25\pm 3$ & NGC 4631 & 12:41:10, 32:45:10 & DBL S & $-4\pm 3$ &          & 22:38:08, 34:17:50 & UNR & $-161\pm 3$\\
         & 10:19:32, 45:37:35 & DBL N & $+7\pm 2$ &          & 12:41:33, 32:46:30 & CMPLX & $-2\pm 3$ & & & & \\
         &                    & 0.8 S & $+8\pm 2$ &          & 12:41:53, 32:27:15 & DBL W FG?& $-38\pm 2$ & & & & \\
         & 10:21:06, 45:23:30 & CMPLX & $+3\pm 3$ &          &                    & 0.6 E FG?& $-28\pm 3$ & & & & \\
         &                    &       &           &          & 12:43:25, 32:25:40 & UNR & $+4\pm 1$ & & & & \\
\hline
\end{tabular}
\end{center}
\end{sidewaystable*}

From the values in Table~\ref{table:fieldRM} it is apparent that when
multiple, double-lobed sources are detected in an individual galaxy
field then they generally have RMs that are in good agreement with one
another. Unresolved sources in the field have RMs which are sometimes
consistent, but seem to have a larger intrinsic scatter. Furthermore,
when the lobes of double-lobed sources are within a factor of two in
brightness they generally have better RM agreement, often as good as
1--2~rad~m$^{-2}$. For more extreme lobe brightness ratios, this
consistency declines. To the extent that a single Galactic
foreground RM contribution is appropriate in a particular field,
the scatter in the measured RMs of the background sources is likely
caused by variations in the intrinsic RMs of the sources themselves.

Since the likely red-shift of the luminous edge-brightened double
sources we detect is greater than about z~=~0.8
\citep{condon_etal_1998}, the associated physical sizes are likely in
the range 0.5--5~Mpc. The large physical separation of such radio
lobes from the host galaxy makes it unlikely that high densities of
thermal electrons will be mixed with the emitting regions. A relevant
phenomenon which has been documented is the tendency for enhanced
Faraday depolarization and RM fluctuations to be seen toward the
fainter lobe of a pair (in both Stokes $I$ and particularly $P$)
\citep{laing_1988, garrington_etal_1988, laing_etal_2006}. This
phenomenon is consistent with the fainter lobe being the more distant
one and its radiation suffering additional propagation effects while
passing through the magneto-ionic halo of the host galaxy. A
prediction of this interpretation is that edge-brightened doubles with
equal brightness lobes are least likely to have differences in their
associated Faraday depth, as we confirm. In any case, it is likely that
well-separated lobes of luminous radio galaxies can provide a good
estimate of the line-of-sight rotation measure with a minimal
intrinsic contribution, in contrast to unresolved and possibly
core-dominated AGN, for which a local host contribution to the Faraday
depth is more likely.

A plausible method of estimating the Galactic foreground contribution
to the RM in each field seems to be a weighted median value
whereby double-lobed sources, particularly those of similar lobe
brightness, are given a very high weight. Some discussion of these
considerations is given below for each target field in turn, with the
result listed in Table~\ref{table:summary}. While the majority of our
target galaxies are well-removed from the Galactic plane
($|b|~>~25^\circ$), where it is plausible that a single foreground RM
may be expected to apply to a region of $34'\times 34'$, the two
exceptions are the fields containing NGC~7331 ($b~=~-21^\circ$) and
NGC6946 ($b~=~+12^\circ$). Diffuse, patchy polarized emission from the
Galaxy is apparent in these fields, accentuating the likelihood that
foreground RM fluctuations may also be present. We stress that we have
in no case made use of polarized emission from the target galaxy
itself to estimate the Galactic foreground RM, since this would bias
the outcome by artificially imposing a zero mean RM on the target
galaxy. For those targets in the direction of the Virgo cluster, or
more generally along the Super-galactic plane, there is also the
possibility that a non-zero contribution to the RM seen toward the
distant background radio galaxies of Table~\ref{table:fieldRM} arises
within these media. Detection of such a contribution would require a
much more extensive sampling of the RM sky, such as envisioned for the
Square Kilometre Array and its Pathfinders
\citep[e.g.][]{johnston_etal_2008}.

\subsection{Notes on Individual Galaxies\label{subsection:notes}}

Here, we discuss the polarized features detected in each of the target
fields. For much of the analysis, we utilize the 18+22cm (deconvolved)
Faraday cubes, and the associated peak $P$ and $\hat{\phi}$ maps shown in
Figures \ref{figure:images:a} and \ref{figure:pkphi}. Comparisons
with optical images are shown in Figure \ref{figure:opt}. We occasionally
refer to the Faraday cubes produced using the 22cm data alone. Trends in
the azimuthal and radial variations in polarized flux and Faraday RM lead
to a generic, consistent picture of the global magnetic field geometry
in the spiral galaxies included in our sample
(see \S\,\ref{subsection:Bdist}). These trends and the global
magnetic field geometry are discussed in detail in Paper III. Most
galaxies with extended polarized flux show signs of broadened Faraday
dispersion functions in small localized regions; see
\S\,\ref{subsection:extprof} for details. All galaxies with compact
(circum-)nuclear polarized emission show signs of significant Faraday
structure in their nuclei. This is discussed in \S\,\ref{subsection:nuc}.

\begin{figure*}
\centering
\resizebox{0.85\hsize}{!}{\includegraphics{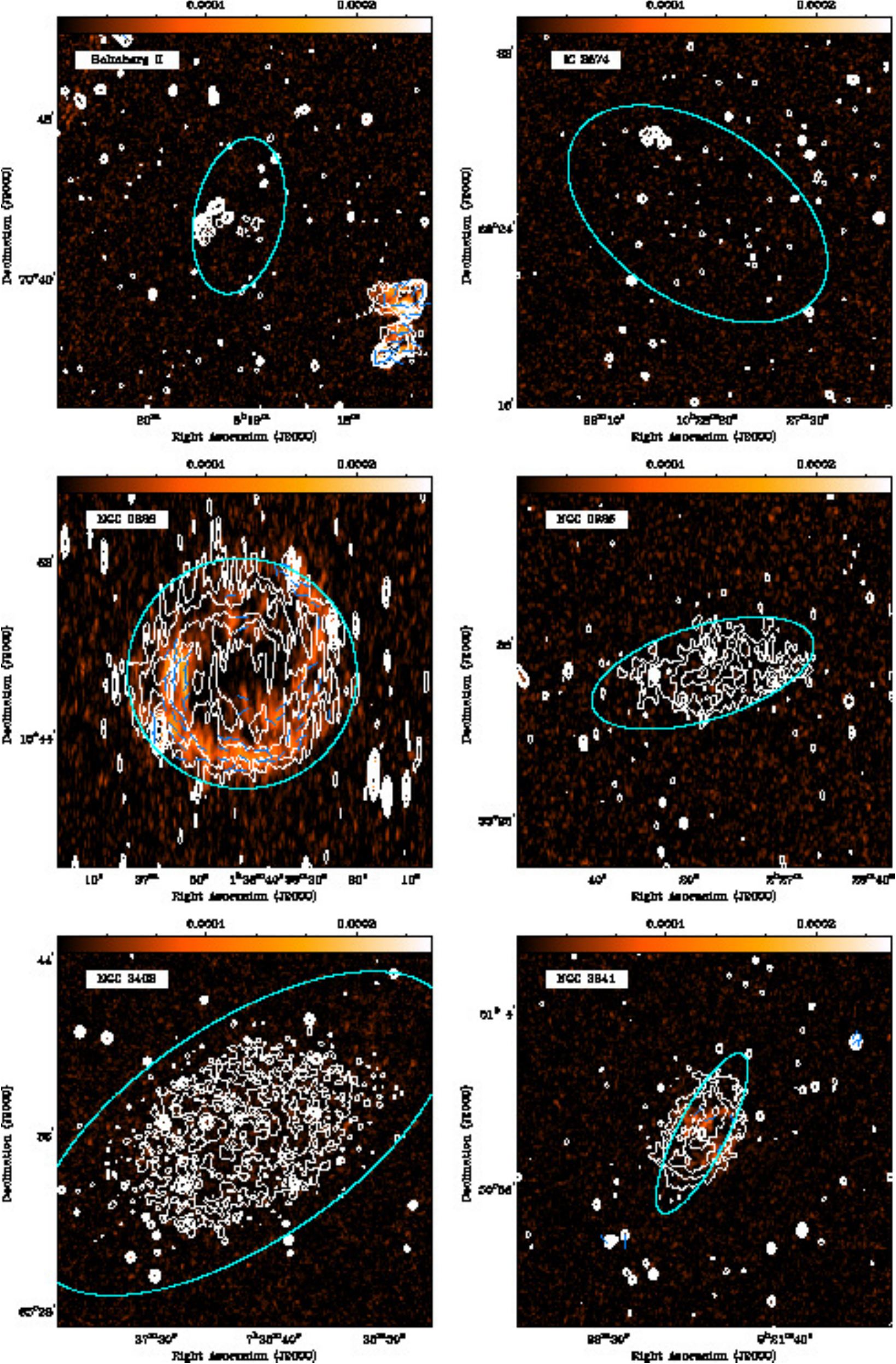}}
\caption{Images of sample galaxies. Contour levels (\emph{white}) in
  Stokes I run from $0.1\,\mathrm{mJy\,beam^{-1}}$ in powers of
  two. The color range of the peak polarized intensity in each
  $F(\phi)$ spectrum is displayed in the colorbar at the top of each
  panel, in units of Jy beam$^{-1}$.
  The galaxy ID is indicated in the upper left of each
  panel. In targets with sufficient polarized emission, the magnetic
  field orientations are displayed with blue vectors. The cyan ellipse
  indicates $D_{25}$.}
\label{figure:images:a}
\end{figure*}

\addtocounter{figure}{-1}
\begin{figure*}
\centering
\resizebox{0.85\hsize}{!}{\includegraphics{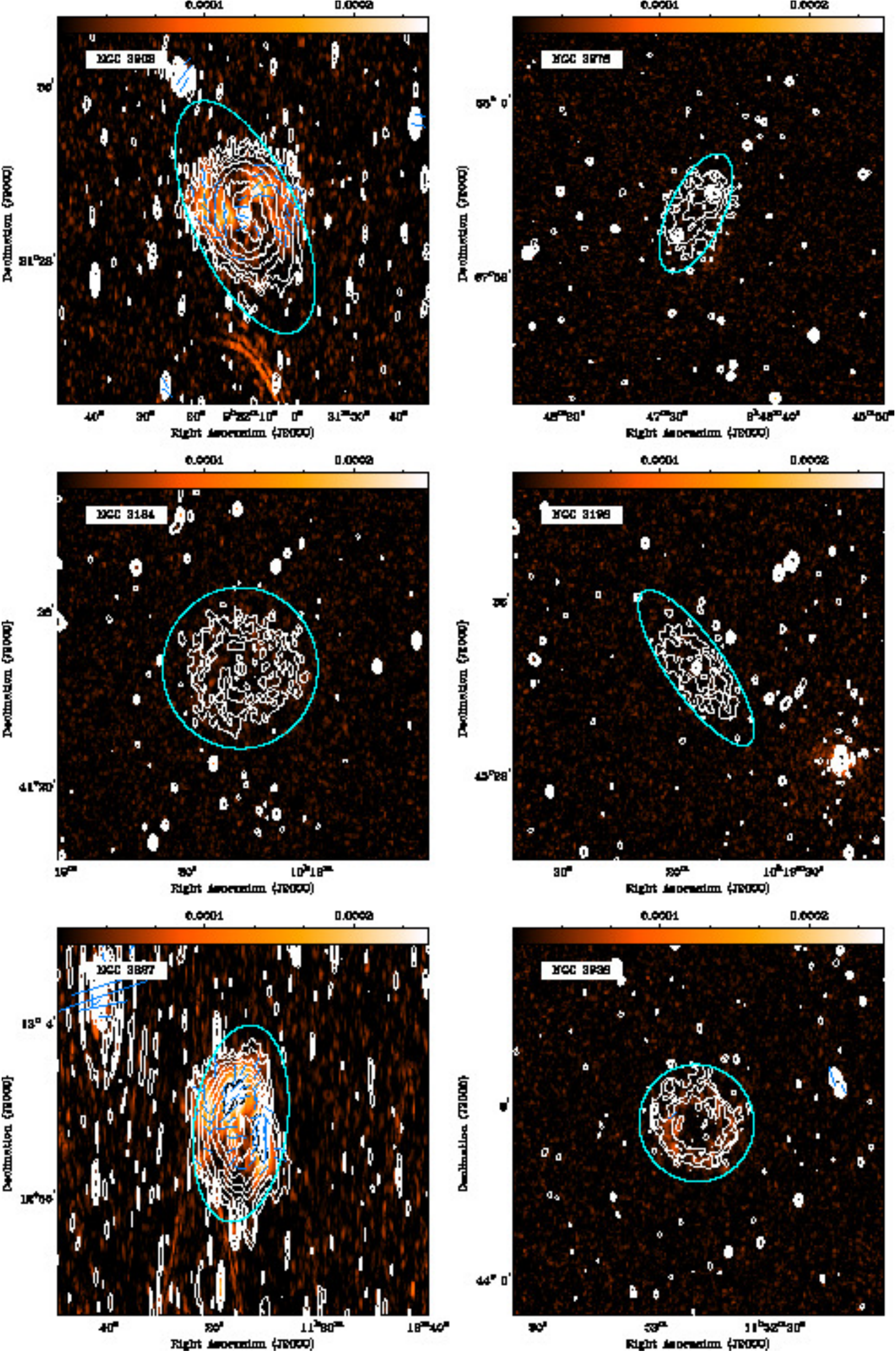}}
\caption{(continued) Images of sample galaxies.}
\end{figure*}

\addtocounter{figure}{-1}
\begin{figure*}
\centering
\resizebox{0.85\hsize}{!}{\includegraphics{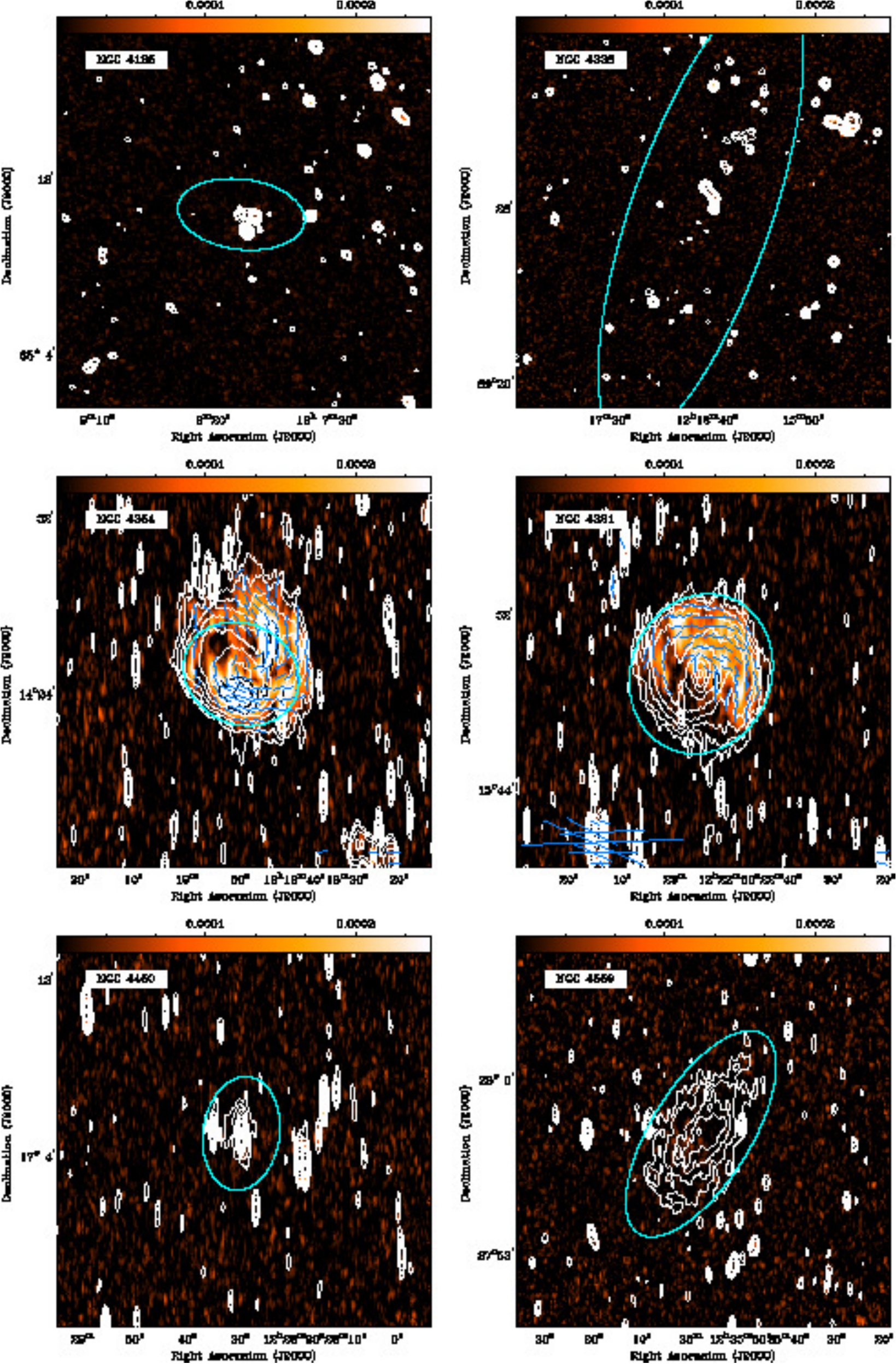}}
\caption{(continued) Images of sample galaxies.}
\end{figure*}

\addtocounter{figure}{-1}
\begin{figure*}
\centering
\resizebox{0.85\hsize}{!}{\includegraphics{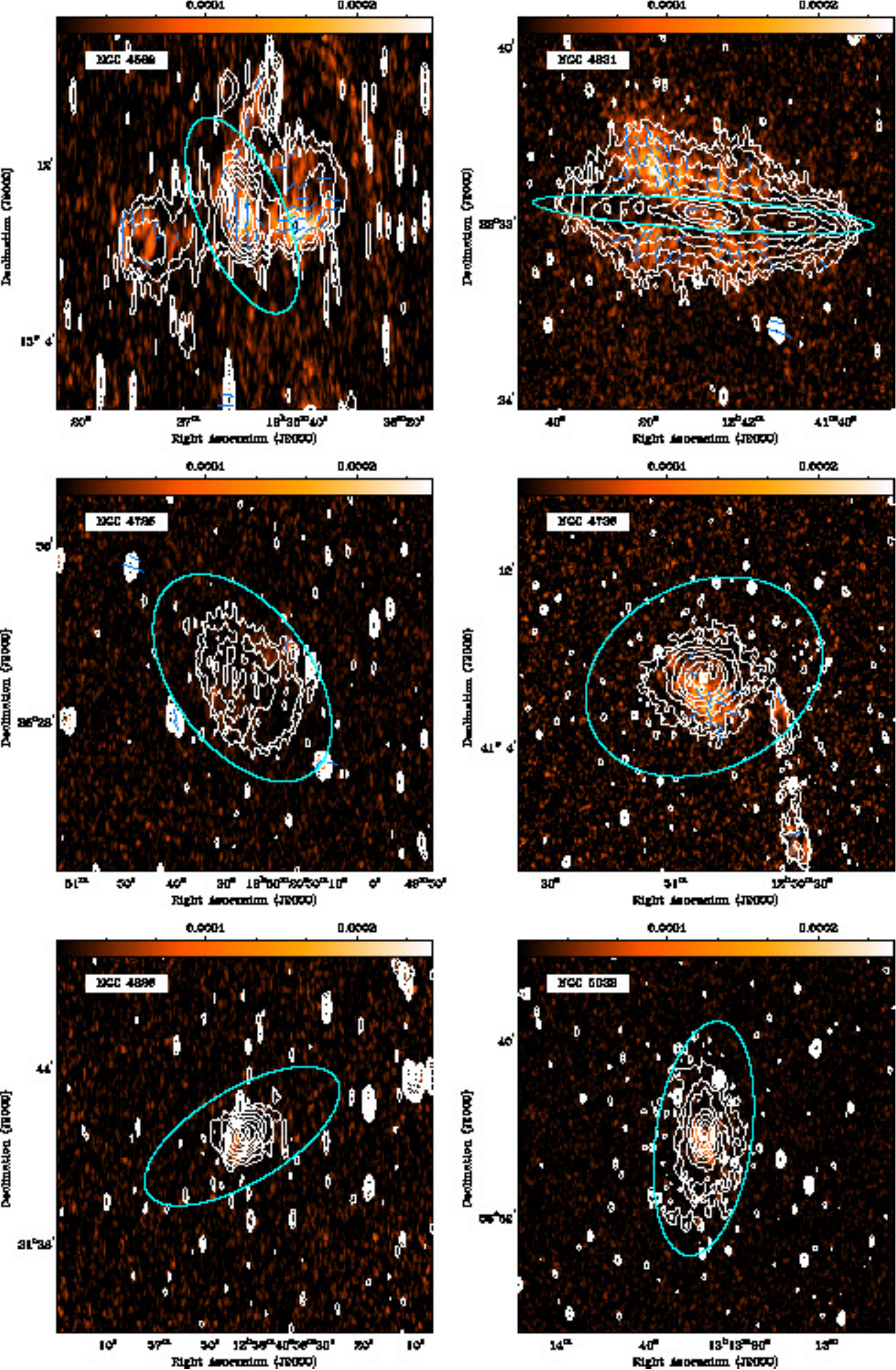}}
\caption{(continued) Images of sample galaxies.}
\end{figure*}

\addtocounter{figure}{-1}
\begin{figure*}
\centering
\resizebox{0.85\hsize}{!}{\includegraphics{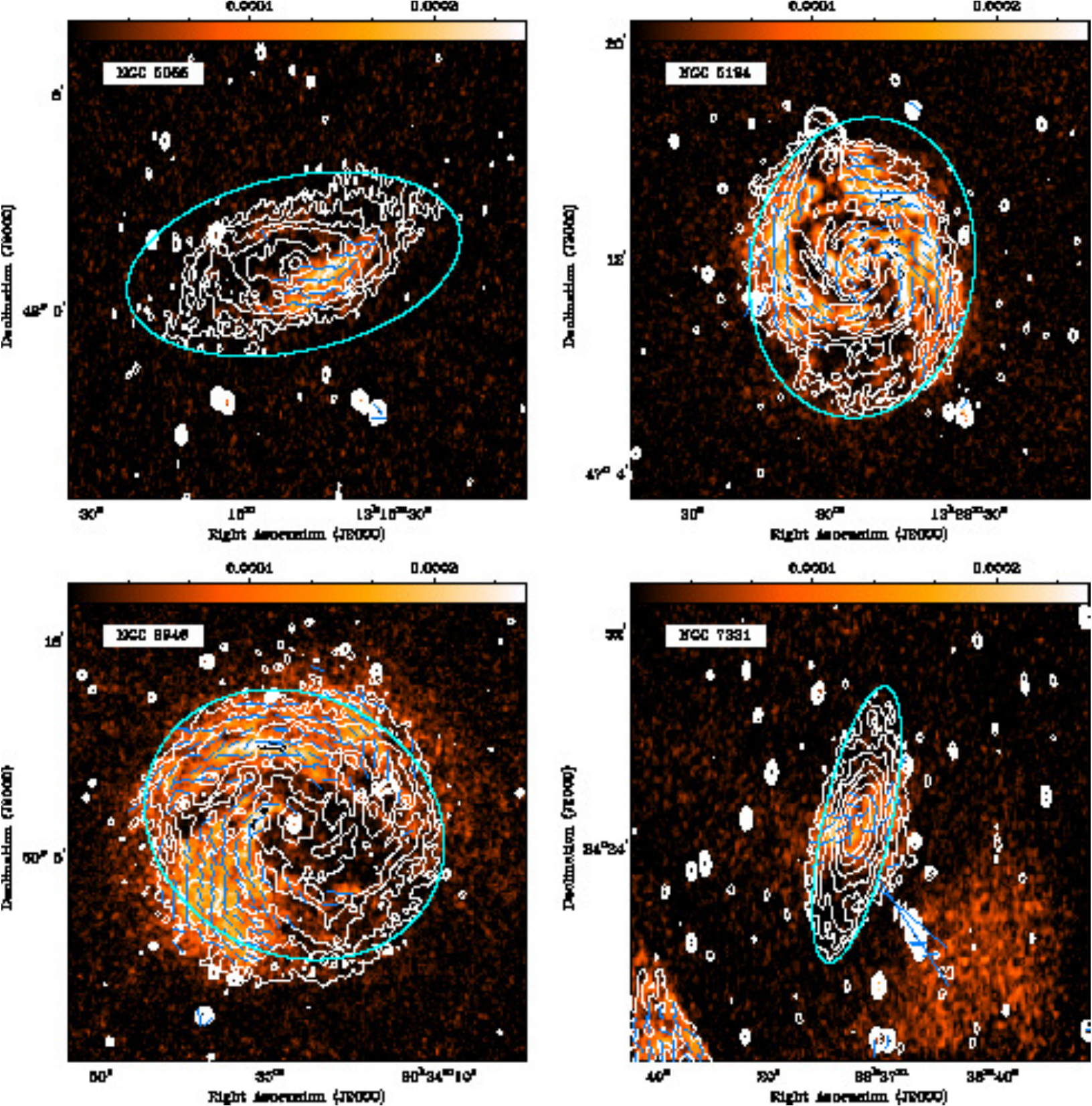}}
\caption{(continued) Images of sample galaxies.}
\end{figure*}

\paragraph{Holmberg II}
There is no convincingly detected polarized emission from Holmberg
II. There is possibly very faint (at/below about the $4\sigma$ level)
emission
at the location of the higher brightness features in the eastern part
of the galaxy which, as discussed by \citet{braun_etal_2007}, are
associated with H$\alpha$-emitting regions. This possible emission is
not seen in the map produced using only the 22cm data. However, to the
southwest (visible in Figure \ref{figure:images:a}) is a classic
double-lobed radio
source which is strongly polarized. The source is catalogued as 6C
B081222.3+704711. The morphology of the polarized emission mostly
fills the region of continuum emission, but with strong depolarization
channels in the northern lobe. The southern lobe is somewhat brighter
in both polarized and unpolarized emission. The polarized fraction is
about 3--4\% in the core, and about 10--20\% along the jet axis. There
are localized regions of higher polarized fraction, at about the 40\%
level, and even reaching as high as 50\%. The magnetic field
orientation is parallel to the northern and southern lobes on their
western edges, but on the eastern edges is perpendicular to the lobes,
where the total continuum morphology also suggests a smooth decline
toward the east. Despite the fact that the two lobes have nearly equal
integrated brightness in $P$, they display RMs
which follow the pattern noted above in
\S\ref{subsection:backgroundRM} for a larger degree of RM fluctuations
to be seen toward the fainter, northern lobe.  Given the obvious
depolarization channels, this difference is perhaps not too
surprising. In view of the 1.5 arcminute angular separation of the
depolarization channels in the northern lobe from the nuclear
position, it would require either a very extended dispersive halo of
the host galaxy, a localized source of significant rotation, or
internal depolarization to
account for both the RM fluctuations and depolarization. Unfortunately
no red-shift information is available for this source, although the
nucleus appears to be coincident with a moderately bright, but
uncatalogued Digital Sky Survey source. The most likely Galactic
foreground RM for this field seems to be about $-10~\pm~2$
rad~m$^{-2}$.

\paragraph{IC 2574}

IC 2574 does not show evidence for significant polarized emission
in either the 22cm map or the 18+22cm map. Since only unresolved
polarized background sources are detected in this field, the Galactic
foreground RM remains quite uncertain at about $-18~\pm~4$
rad~m$^{-2}$. 

\paragraph{NGC 628 (M74)}

This relatively face-on spiral galaxy shows substantial polarized
emission in the form of an incomplete ring near the edge of the
optical disk which is brightest at PA $\sim\,80^\circ-340^\circ$ (Position
Angles measured east of north). The minimum in polarized intensity
occurs at the PA of the receding major axis (PA~=~25$^\circ$, as
tabulated in Table~\ref{table:summary}).  The brightest emission is
associated with two inter-arm regions in the outer galaxy; one
extending from PA $\sim$ 80$^\circ$ to 210$^\circ$ and the other from PA
$\sim$ 200$^\circ$ to 340$^\circ$. These seem to be continuations into the
outer disk of the inner disk interarm regions. The polarized fraction
at 22cm in these regions is approximately 10--20\% at small radii;
increasing to 40--50\% at the largest radii, indicating an
exceptionally well-ordered magnetic field. The magnetic field vectors
are closely aligned with the features themselves. The Faraday depth
distribution shows some systematic variation with PA which we will
discuss in Paper III. Based on the brighter lobes of
the three polarized double sources detected in this field, the likely
Galactic foreground RM is about $-34~\pm~2$ rad~m$^{-2}$.

\paragraph{NGC 925}

There is no polarized emission detected in this galaxy. However,
several faint background sources are seen toward the edges of the
field. Excluding the unresolved polarized source most discrepant from
the single double source lobe, we obtain an estimate of the Galactic
foreground RM with a value of about $-10~\pm~2$ rad~m$^{-2}$.

\paragraph{NGC 2403}

There is a faint polarized component which is predominantly diffuse,
but is too faint to characterize well with the current
observations. Smoothing to a beam size of
$45^{\prime\prime}\times45^{\prime\prime}$
enhances the signal to
noise ratio of some of the polarized emission, which is concentrated in the
western half of the galaxy, with some localized enhancements at the
eastern and western edges of the optical disk. But even after application of
smoothing, it is still extremely faint (typical surface brightness
$\approx\,30-50\,\mu$Jy/beam, reaching as high as
$\approx\,100\,\mu$Jy/beam). The lowest brightness of polarized
emission occurs on the receding major axis (PA~=~125$^\circ$, as
tabulated in Table~\ref{table:summary}). Deeper observations would be
required to further constrain the magnetic field properties in this
target.  The three unresolved polarized sources which are detected in
the field do not permit a good estimation of the Galactic foreground
RM in view of their large scatter.

\paragraph{NGC 2841}

In Stokes $I$, \citet{braun_etal_2007} note a diffuse ``hourglass''
structure, with the long axis of the hourglass oriented perpendicular
to the disk major axis. In the polarized emission, the highest
brightnesses are seen along the minor axis trailing away slowly to the
northwest and more rapidly to the southeast. The lowest brightness of
polarized emission occurs near the receding major axis
(PA~=~153$^\circ$, as tabulated in Table~\ref{table:summary}). The
directions of the magnetic field vectors are primarily aligned along
the polarized arcs.  The three polarized background doubles in this
field display substantial differences in their RMs. The greatest
consistency is seen between the two lobes of one of the doubles,
at $-6~\pm~3$ rad~m$^{-2}$. There may be significant structure
of the foreground RM in this field.

\paragraph{NGC 2903}

In this galaxy, one of the Starburst supplement to the basic SINGS
sample, bright polarized arcs are detected along both sides of the
minor axis trailing away in brightness slowly to the northeast and
more rapidly to the southwest. The lowest brightness of polarized
emission occurs on the receding major axis (PA~=~204$^\circ$, as
tabulated in Table~\ref{table:summary}). The polarized fraction
increases from about $1$\% in the inner parts to about 5--15\% at
intermediate radii, to as high as 40\%. The magnetic field vectors are
roughly parallel to optical spiral arm structures for the minor axis
features, although generally with a slightly larger radial
component. Field lines run almost perpendicular to the linear major
axis feature. The Faraday depth distribution shows a small systematic
variation with PA in the minor axis features which we will comment on
in Paper III, together with a large systematic offset of these (by 60
rad~m$^{-2}$) relative to the major axis feature. Very good
consistency is found for the RMs toward both lobes of a double source
in the field, suggesting a value of $+3~\pm~1$ rad~m$^{-2}$ for the
Galaxy in this direction.

\paragraph{NGC 2976}

There is very faint diffuse polarized emission associated with this galaxy,
together with a modest enhancement along the southwestern edge of the
optical disk. Only a single lobe of one double source is detected in
this field, yielding a Galactic foreground RM of $-34~\pm~2$
rad~m$^{-2}$. The unresolved sources in the field show significant
scatter about this value.

\paragraph{NGC 3184}

In this galaxy, faint polarized emission is apparent over much of the
northern half of the disk. The lowest brightness of polarized emission
occurs near the receding major axis (PA~=~179$^\circ$, as tabulated in
Table~\ref{table:summary}). Interarm regions may be enhanced relative
to the spiral arms, but deeper observations would be required for a
definitive analysis. Only a single lobe of one double source is
detected in this field, yielding a Galactic foreground RM of
$+19~\pm~2$ rad~m$^{-2}$. The unresolved sources in the field show
significant scatter about this value.

\paragraph{NGC 3198}

No significant polarized emission is detected in this target, apart
from a possible detection of the nucleus. An equal brightness double
source in the field allows a very consistent assessment of the
Galactic foreground RM of $+7~\pm~2$ rad~m$^{-2}$.

\paragraph{NGC 3627 (M66)}

Bright polarized emission is detected this galaxy, with a conspicuous
north-south gradient of the fractional polarization. It originates in
both optical spiral arm and in interarm regions. The polarized
fraction at 22cm is less than 1\% in the bright optical bar and inner
disk of the galaxy, and increases in the regions outside of the spiral
arms to as much as 15\% on the eastern minor axis. The lowest
brightness of polarized emission occurs near the receding major axis
(PA~=~173$^\circ$, as tabulated in Table~\ref{table:summary}). The
magnetic field orientation closely follows the optical spiral
structure, except at the largest radii where it generally becomes more
radial. Some possible systematic variation in the Faraday depth can be
discerned which we will comment on in Paper III. Note that
\citet{soida_etal_2001} have published VLA and Effelsberg observations
of this galaxy at 4.8 and 8.5~GHz which detect many of the same
trends, although not detecting the same northern extent. The brighter
lobe of a background triple source permits assessment of the Galactic
foreground RM of $+13~\pm~1$ rad~m$^{-2}$. Other field sources are
scattered around this value.

\paragraph{NGC 3938}

Moderately faint polarized emission is detected in this almost face-on
spiral. The polarized emission is concentrated to the outer disk and
is enhanced in inter-arm regions. In contrast to other galaxies in our
sample, the lowest brightness of polarized emission seem to occur near
the \emph{approaching} rather than the receding major axis (i.e.,
opposite to PA~=~204$^\circ$, as tabulated in
Table~\ref{table:summary}). However, as apparent in the Stokes $I$
imaging \citep{braun_etal_2007}, there is also a minimum in the total
intensity on the approaching major axis, so that this object does not
represent a counter-example to the trend we see in the azimuthal
modulation of $P$. The polarized fraction (where detected) is about
10-25\%. In the few regions where there is enough signal to determine
the magnetic field orientation, the field lines run parallel to the
spiral arms. The brighter lobe of a double source in the field permits
assessment of the Galactic foreground RM of $+1~\pm~2$
rad~m$^{-2}$. Other field sources display only a small scatter around
this value.

\paragraph{NGC 4125}

In this elliptical galaxy, \citet{braun_etal_2007} report a continuum
source at the nucleus, which we find is not polarized. They also report a
double radio galaxy just to the southwest, the southern component of
which is polarized at about the 5\% level. The three double sources in
this field provide a very consistent measurement of the
Galactic foreground RM of $+19~\pm~1$ rad~m$^{-2}$. 

\paragraph{NGC 4236}

In this galaxy, \citet{braun_etal_2007} report detecting continuum
emission from the bright knots in the disk. There is no polarized
counterpart associated with these features. The double background
radio source behind the disk of NGC~4236 is detected in polarization,
at about the 7\% polarized fraction level with an RM of $-3~\pm~2$
rad~m$^{-2}$. The two unconfused background double sources in the
field provide a consistent estimate of the Galactic foreground RM of
$+15~\pm~2$ rad~m$^{-2}$. The difference may well be due to the
magneto-ionic ISM of NGC~4236.

\begin{figure*}
\centering
\resizebox{0.85\hsize}{!}{\includegraphics{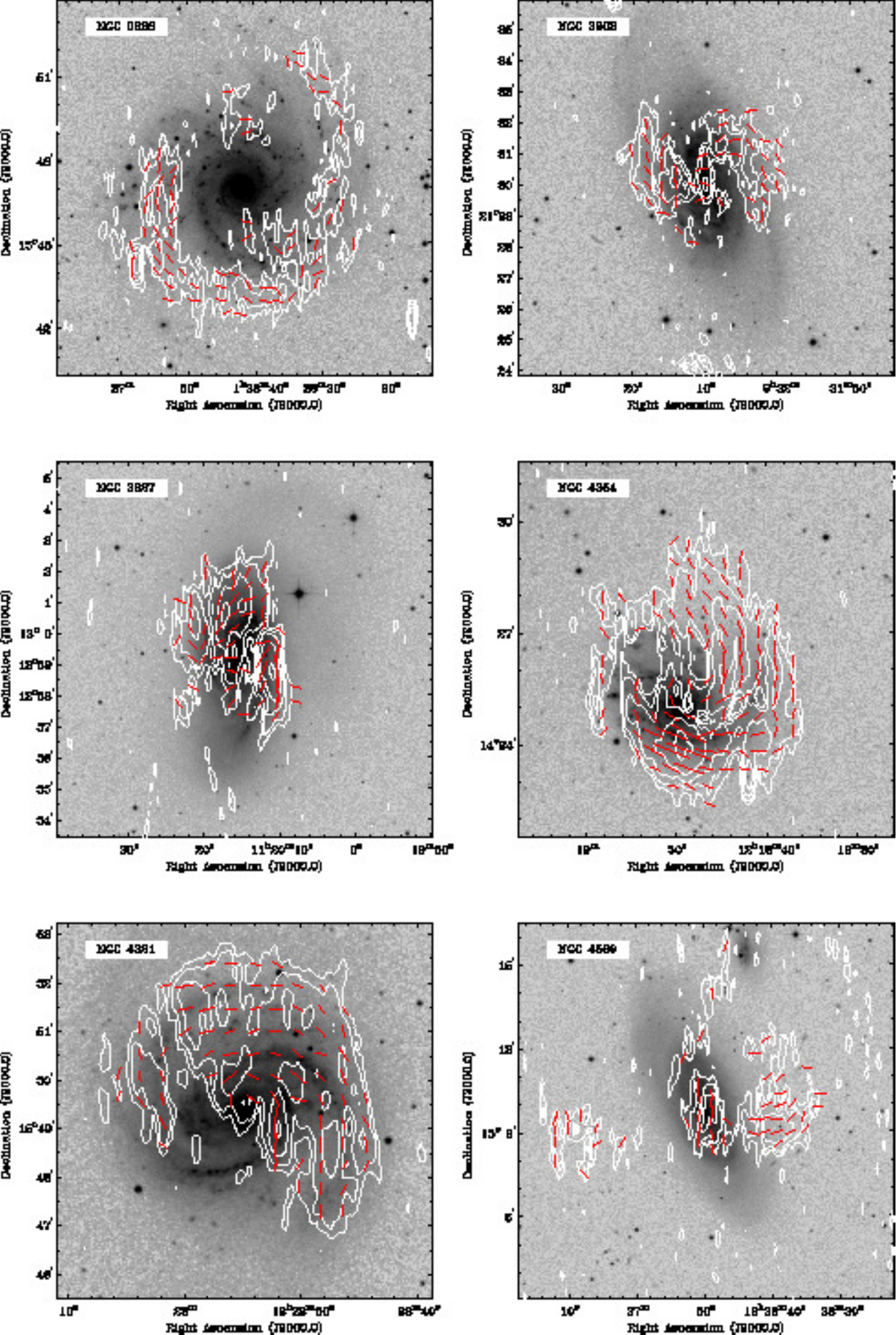}}
\caption{Optical images of sample galaxies with extended polarized
  flux. The background images are red plates from the DSS-2, and are
  presented with a square-root color transfer to bring out the faint
  structures. Contour levels (\emph{white}) of polarized intensity run
  from $50\,\mu\mathrm{Jy\,beam^{-1}}$ in powers of 2.
  The galaxy ID is indicated in the upper left of each panel. The
  magnetic field orientations are displayed with red vectors.}
\label{figure:opt}
\end{figure*}

\addtocounter{figure}{-1}
\begin{figure*}
\centering
\resizebox{0.85\hsize}{!}{\includegraphics{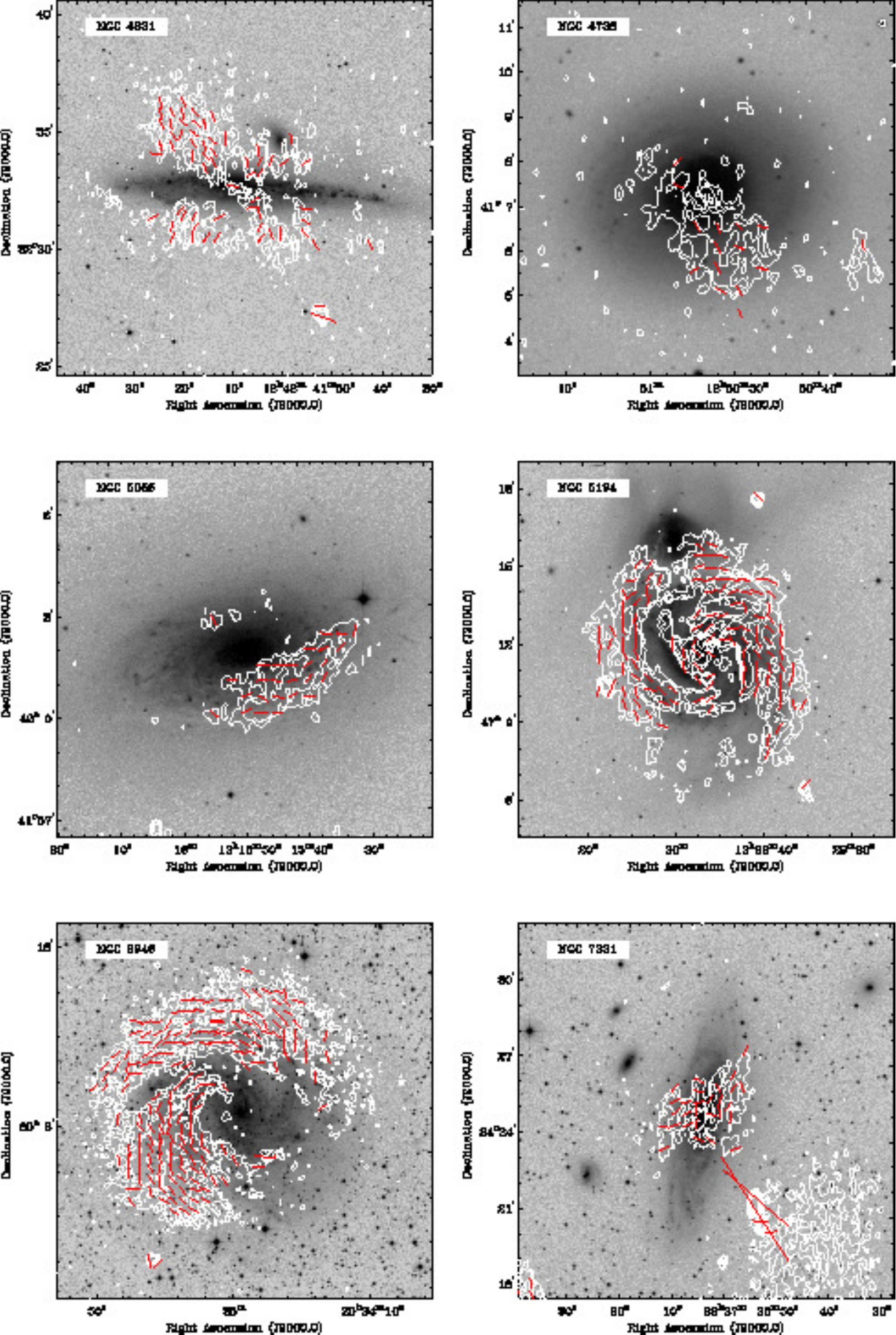}}
\caption{(continued) Optical images of sample galaxies with extended
  polarized flux.}
\end{figure*}

\paragraph{NGC 4254 (M99)}

Bright polarized emission is detected from an incomplete ring
extending from PA $\sim$ 90 -- 330$^\circ$. The peak of the polarized
continuum is to the south of the southern spiral arm. The Stokes $I$ peak is
also slightly offset to the south from the nucleus of the galaxy, though not
as far as the peak of the polarized emission. The lowest brightness of
polarized emission occurs near the receding major axis
(PA~=~65$^\circ$, as tabulated in Table~\ref{table:summary}). Both the
interior and exterior of the unusual western spiral arm display bright
polarized emission that subsequently extends far to the north of the
optical disk. The polarized fraction is only moderate in the vicinity
of the southern peak; 4\% at the peak and 1--10\% elsewhere in that
region. Polarized fractions are also low in the inner spiral arm
region; below one percent ranging up to a few percent. Along the outer
western edge and in the northern polarized extension, the polarized
fraction is generally in the range 10--15\%, but in some areas as high
as 20--25\%. The magnetic field lines follow the spiral arm structure
very well, especially in the southern region where the polarized
emission is brightest. To the north, at the location of the radio continuum
extension, the magnetic fields continue to follow the direction
defined by the optical spiral arm structure, even though the optical
arm is no longer detected. The apparent departure from this simple
pattern in the southwest is due to a polarized background source.  The
Faraday depth distribution shows some interesting systematic
variation: the Faraday depth tends to be negative on the outside
of the spiral arms, and positive on the inside.
Within the bright radio continuum disk there may be
evidence for some azimuthal variation, while the western and northern
polarized extensions do not participate in this pattern. This galaxy
has recently been studied by \citet{chyzy_2007} and
\citet{chyzy_2008}, who report on VLA and Effelsberg polarimetric
observations at 1.4, 4.8 and 8.5~GHz. A single double source in the
field allows reasonable assessment of the Galactic foreground RM (from
the brighter lobe) of $-13~\pm~3$ rad~m$^{-2}$.

\paragraph{NGC 4321 (M100)}

In this spiral galaxy, polarized emission is clearly detected
throughout most of the disk. In the northwest, bright polarized
emission is present throughout the interarm regions. But in the
southeast the polarized surface brigntess declines dramatically. The
lowest brightness of polarized emission occurs near the receding major
axis (PA~=~159$^\circ$, as tabulated in
Table~\ref{table:summary}). The polarized fraction in this disk is
generally rather low, ranging from less than or about 2\% in the inner
regions to about 5--10\% in the interarm regions, with the polarized
fraction tending to be higher outside of the arm than inside the
arm. At the edges, the polarized fraction is higher at about the 15\%
level, and localized spots where the fraction reaches 30\%. The
appearance of the magnetic field lines is highly ordered, and follows
the orientations of the optical spiral arms. An almost equal double in
the field allows consistent assessment of the Galactic foreground RM
of $-17~\pm~1$ rad~m$^{-2}$.

\paragraph{NGC 4450}

No significant polarized emission is detected in this galaxy. The
almost equal double and brighter lobe of a triple in the field allow
consistent assessment of the Galactic foreground RM of $-8~\pm~1$
rad~m$^{-2}$.

\paragraph{NGC 4559}

In this spiral galaxy, diffuse continuum emission is detected, but the
polarized component is extremely faint and only detected in a small
region in the southeast portion of the disk. This asymmetry is again
consistent with the lowest polarized intensity to be seen on the
receding major axis. Deeper observations would
be needed to better characterize the polarized emission. The equal
double in the southern part of the field provides a consistent
estimate of the Galactic foreground RM of $-5~\pm~2$ rad~m$^{-2}$,
while the remaining four unresolved sources in the north and east of
the field are all tightly clustered around an RM of $+6~\pm~2$
rad~m$^{-2}$. This systematic difference of the RM by of $\sim 10$
rad~m$^{-2}$ over $\sim15'$ in a field so near the
Galactic pole ($b~=~86^\circ$) is surprising.

\paragraph{NGC 4569 (M90)}

In this moderately inclined spiral galaxy, polarized emission is
detected in the central disk region on either side of the minor
axis. Polarized intensity declines more slowly to the southwest and
more rapidly to the northeast (where the receding major axis is
located (PA~=~23$^\circ$, as tabulated in Table~\ref{table:summary}).
Moreover, the spectacular double lobe extension which is oriented
roughly along the minor axis is also detected in polarization,
particularly along its edges. Even more interesting is the continuum
bridge connecting the galaxy to its small companion, IC 3583 
(located about 6 arcminutes to the northwest, and visible in both
Figures \ref{figure:images:a} and \ref{figure:opt}), which
also has a polarized counterpart. The large-scale structures in
our map have already been observed at lower spatial resolution and
analyzed by \citet{chyzy_etal_2006}, who observed this system with
the Effelsberg telescope. The polarized bridge and the lobe
extensions have rather high polarized fractions. The bridge is
polarized at the 20--30\% level, the lobes at the 10-20\% level, with
localized hot-spots of higher polarized fraction, of about 40\%. The
disk itself has polarized fractions of only about 1-2\%. As for the
magnetic field orientations, the situation is confused in the disk due
to the modest angular resolution, but there seems to be a slight
tendency for field lines to follow the optical spiral arms. In the
radio lobes, the field lines appear to trace the edges of apparent
cavities. Finally, in the extension toward IC 3583, the magnetic field
lines run roughly along the direction of the extension. The brighter
lobe of a background double source provides an estimate of the
Galactic foreground RM in this direction of $+18~\pm~2$ rad~m$^{-2}$.

\paragraph{NGC 4631}

In this edge-on interacting spiral, the polarized emission
is found in a roughly X-shaped morphology, and comes mainly from the
extraplanar regions. The disk itself seems to be largely
depolarized. Polarized emission is detected in the central region and
in each of the four extraplanar galaxy quadrants. The north side is
brighter in polarization than the south side. The brightest polarized
intensity is from the northeast quadrant, which is the region where the
dramatic HI extension studied by \citet{rand_1994} is located. The polarized
structure runs roughly parallel with the HI extension, but fills the
region between the disk and the HI filament. The polarized fraction in
this galaxy is less than one percent in the central regions and
increases with height above the midplane. At the largest $z$-heights,
the polarized fraction reaches as high as 30--40\% in some places. The
magnetic field lines run along the X-shaped polarized morphology. In
the northeast quadrant, they run almost parallel to the HI extension,
but these seem to be unrelated. The polarized structures reported
here have been observed previously by \citet{hummel_etal_1991} and
\citet{golla_hummel_1994}. The best estimate of the Galatic
foreground RM in this direction comes from the unconfused double
source in the field with an RM of $-4~\pm~3$ rad~m$^{-2}$; consistent
with several other sources in the field. The double
background source lying just south of NGC~4631 is likely to be
strongly affected by the halo of that galaxy, the brighter lobe of
which displays an RM of $-38~\pm~2$ rad~m$^{-2}$.

\paragraph{NGC 4725}

Extremely faint polarized emission is detected in this moderately
inclined barred spiral galaxy, with the polarized emission originating
at both ends of the minor axis. The polarized emission avoids the bar,
which is at a position angle of about 45 degrees, and is mostly found
on the outer periphery of the ring-like structure. The polarized
emission is too faint to allow investigation of its detailed
properties; deeper observations would be required. It is not possible to say
anything about the magnetic field orientation, as too little signal is
available. The foreground RM from the Galaxy in this direction can be
estimated from the double radio source in the field at $+4~\pm~4$
rad~m$^{-2}$; a value consistent with several other unresolved sources
in the field.

\paragraph{NGC 4736 (M94)}

The polarized emission in this galaxy is seen from both the inner
star-forming disk and also concentrated along the minor axis,
particularly in the form of a possible polarized lobe directed toward
the southwest. Within the central disk, the polarized intensity
declines to a minimum in the direction of the receding major axis
(PA~=~296$^\circ$, as tabulated in Table~\ref{table:summary}). The
central source is polarized at about the 2\% level. The inner ring is
polarized on the south side, at about the same fraction. The
polarized fraction increases at large radii up to about 40\%. Magnetic
field lines in the possible lobe structure are aligned radially away
from the nucleus. Since the near-side of the stellar disk in this
system (as determined from optical dust lanes) is in the northeast,
the location of the southwestern lobe is consistent with it being the
closer of a pair of symmetric nuclear outflows, in which the more
distant lobe suffers greater depolarization from the intervening
disk. This, together with the lack of a conspicuously distinct
feature in the Stokes $I$ map at the same location, points to the lobe
structure being intrinsic to NGC 4736 (as opposed to an extended, polarized,
background source). Galactic foreground RM in the field can be estimated
from several double sources which are detected in polarization. Two of
these are rather faint and have large uncertainty, while the high
signal-to-noise detection yields a value of $+1~\pm~1$ rad~m$^{-2}$. A
very extended (6 arcmin) background double radio galaxy is also seen in the
southwest portion of the field. Although both lobes have similar
brightness in both $I$ and $P$, the polarized surface brightness is so
faint that an accurate RM determination is not practical.

\paragraph{NGC 4826 (M64)}

There is a low level of polarized emission detected in the southern
quadrant of this system
that nowhere exceeds 4$\sigma$. The best estimate of the Galactic
foreground RM in this field comes from the brighter lobe of a double
radio source yielding $-8~\pm~2$ rad~m$^{-2}$.

\paragraph{NGC 5033}

The polarized emission in this galaxy is associated with the bright
inner continuum disk reported by \citet{braun_etal_2007}. It has a
roughly X-shaped appearance, which may be indicative of minor axis
outflows, as seen elsewhere in our sample. The polarized brightness
declines to a minimum in the direction of the receding major axis
(PA~=~352$^\circ$, as tabulated in Table~\ref{table:summary}). In the
central parts the polarized fraction is of the order of
$\lesssim\,1$\%. At larger radii the polarized fraction increases to
about 5-7\%. Although there are no well resolved double radio sources
in the field, at least one source is observed to be somewhat extended
with an RM of $+9~\pm~2$ rad~m$^{-2}$. The various other unresolved
sources in the field show scatter about this value.

\paragraph{NGC 5055 (M63)}

Diffuse polarized emission is detected from the disk of this inclined
galaxy on both sides of the minor axis. The highest brightnesses are
associated with the zone of strong warping of the gaseous disk in the
southwest at the edge of the star-forming disk. A fainter counterpart
is seen in the northeast. The minimum in polarized intensity occurs at
the PA of the receding major axis (PA~=~102$^\circ$, as tabulated in
Table~\ref{table:summary}). The polarized fraction of the brightest
feature is mainly in the range of 5--10\%, but at the southernmost end
(where the contribution from the bright inner disk is significantly
fainter) the fraction increases to 15--25\%. The magnetic field lines
closely follow the spiral arm structure observed in the optical
image. Two background double sources yield an estimate of the
background RM in this field of about $-8~\pm~3$ rad~m$^{-2}$.

\paragraph{NGC 5194 (M51)}

In M51, the polarized emission is clearly detected throughout the
disk, though there are large variations in the polarized fraction. The
bright polarized emission traces out a spiral pattern that runs
parallel to the optical arms. The minimum in polarized intensity occurs
at the PA of the receding major axis (PA~=~172$^\circ$, as tabulated in
Table~\ref{table:summary}). The companion, NGC 5195, is not
detected in polarization. The polarization fraction at 22cm in M51 is
variable, remaining lower than 5\% within most of the optical disk,
and increasing at large radii, beyond the outer spiral arms, to as
much as 25--30\%. The orientation of the magnetic field lines closely
tracks both the large-scale spiral pattern
\citep[as also seen by][see their Figure 10]{horellou_etal_1992},
as well as small-scale
features that often have dust-lane counterparts in the optical
imagery. In two locations on the eastern side of the disk, the
polarized emission crosses from the inside of the optical arm to the
outside; at the crossing point, the magnetic field vectors turn from
running parallel to the spiral to follow the polarized emission across
the arm. These features occur near
$(\alpha_{\mathrm{J2000}},\delta_{\mathrm{J2000}})\,=\,$
(13:30:1.5,47:12:15) and (13:30:5.4,47:10:30).
Systematic variation in the Faraday depth is seen as
function of azimuth, which will be discussed in Paper III. Two
background double radio sources are detected with very high
signal-to-noise, including one with comparable integrated $P$ from
each lobe. In Stokes $I$, the brightness ratio of the two lobes in
that source is actually 0.75, with the western lobe being the
brighter. The best estimate of the Galactic foreground RM for this
field is $+12~\pm~2$ rad~m$^{-2}$.

\paragraph{NGC 6946}

The polarized emission is strongly detected from the northeast half of
this galaxy, tapering away toward the southwest. The minimum in
polarized intensity occurs at the PA of the receding major axis
(PA~=~243$^\circ$, as tabulated in Table~\ref{table:summary}). The
polarized fraction (in the northeast) is quite low in the inner parts
at less than 10\%, moderate at intermediate radii at about 20--30\%,
and very high in the outer parts, reaching up to (and perhaps above)
40--50\%. The magnetic field lines are closely related to the
large-scale spiral morphology traced by massive star formation and
dust lanes, running largely parallel to the optical arms.
As has been pointed out by \citet{beck_2007}, the
peaks in polarized emission originate in the interarm regions.
The Faraday depth distribution shows a very clear
systematic variation with azimuth, and will be discussed in Paper III.
\citet{beck_2007} have recently reported VLA and Effelsberg
observations of this galaxy at 1.4, 2.6, 4.8, 8.5 and 10.6 GHz. They
demonstrate that there is substantial depolarization at 20cm in the
southwest quadrant relative to higher frequencies. The large-scale
features that he discusses are very similar to this work.

Unfortunately
only a single well-resolved double background source is detected in
this field (about $10^{\prime}$ southwest of NGC~6946), and this source is quite
asymmetric. The brighter lobe has a well-defined RM of $-14~\pm~2$
rad~m$^{-2}$. Several other sources in the field are possibly
influenced by RM contributions from the disk of NGC~6946 itself;
including the extended background source at
($\alpha$ 20:35:19, $\delta$ +60:02:05) just $2^{\prime}$
south of the NGC~6946 disk that appears to be a barely resolved
($30^{\prime\prime}$)
double with a well-defined RM of $+23~\pm~2$ rad~m$^{-2}$. As
previously noted, the low Galactic latitude of this field
($b~=~+12^\circ$, decreasing to the southeast) enhances the likelihood
of fluctuations in the foreground RM. We suggest that the most likely
value of the foreground affecting NGC~6946 is an RM of $+23~\pm~2$
rad~m$^{-2}$, but stress that there is a substantial systematic
uncertainty in this value. \citet{ehle_beck_1993} and
\citet{beck_2007} have previously determined a value of
$\rpms{\approx\,+40}$ in this field, which is consistent with the
mean Faraday depth that we have observed in the disk of NGC~6946
(see Figure \ref{figure:pkphi}). We note that their foreground RM
value is derived using the diffuse emission of NGC~6946 itself, while
our derivation was performed using only background sources in the field.
We postulate that the difference may be due to a non-zero contribution
to the rotation measures in NGC~6946 from a vertical component of the
magnetic field in the halo of that galaxy. We return to this possibility
in Paper III.

\paragraph{NGC 7331}

The bright inner disk of this highly inclined spiral disk is
highly polarized, and polarization is also detected in the outer disk
on both sides of the minor axis, extending well out into
the low surface brightness outer disk. The polarized fraction is quite
low in the inner parts, at about 1--3\%, but it increases rapidly
toward the outer parts up to about 20--40\%. This is a galaxy in which
multiple Faraday depth components are encountered along some
lines-of-sight, implying large-scale interspersal of emitting and
rotating media. Magnetic field orientations of the single brightest
polarized component along each line-of-sight show some tendency to be
aligned with the star-forming disk at small radii, but become
increasingly radial in the minor axis extensions. The peak Faraday depth
distribution is bi-modal, with two dominant ranges occurring, one near
$\phi~=~0$ rad~m$^{-2}$ in the inner disk and the other near $-150$
rad~m$^{-2}$ associated with the minor axis structures. The Galactic
foreground RM in this general direction is known to be quite extreme
\citep[cf.][]{broten_etal_1988, han_qiao_1994}, even at tens of
degrees from the Galactic plane. The brighter lobes of two resolved double
sources in the field suggest a value of about $-177~\pm~7$
rad~m$^{-2}$. As was the case for NGC~6946, the relatively low
Galactic latitude ($b~=~-21^\circ$) increases the
likelihood of fluctuations in the foreground RM. This is reflected in
the larger scatter of RM values of even the most reliable of probes.

\begin{figure*}
\resizebox{\hsize}{!}{\includegraphics{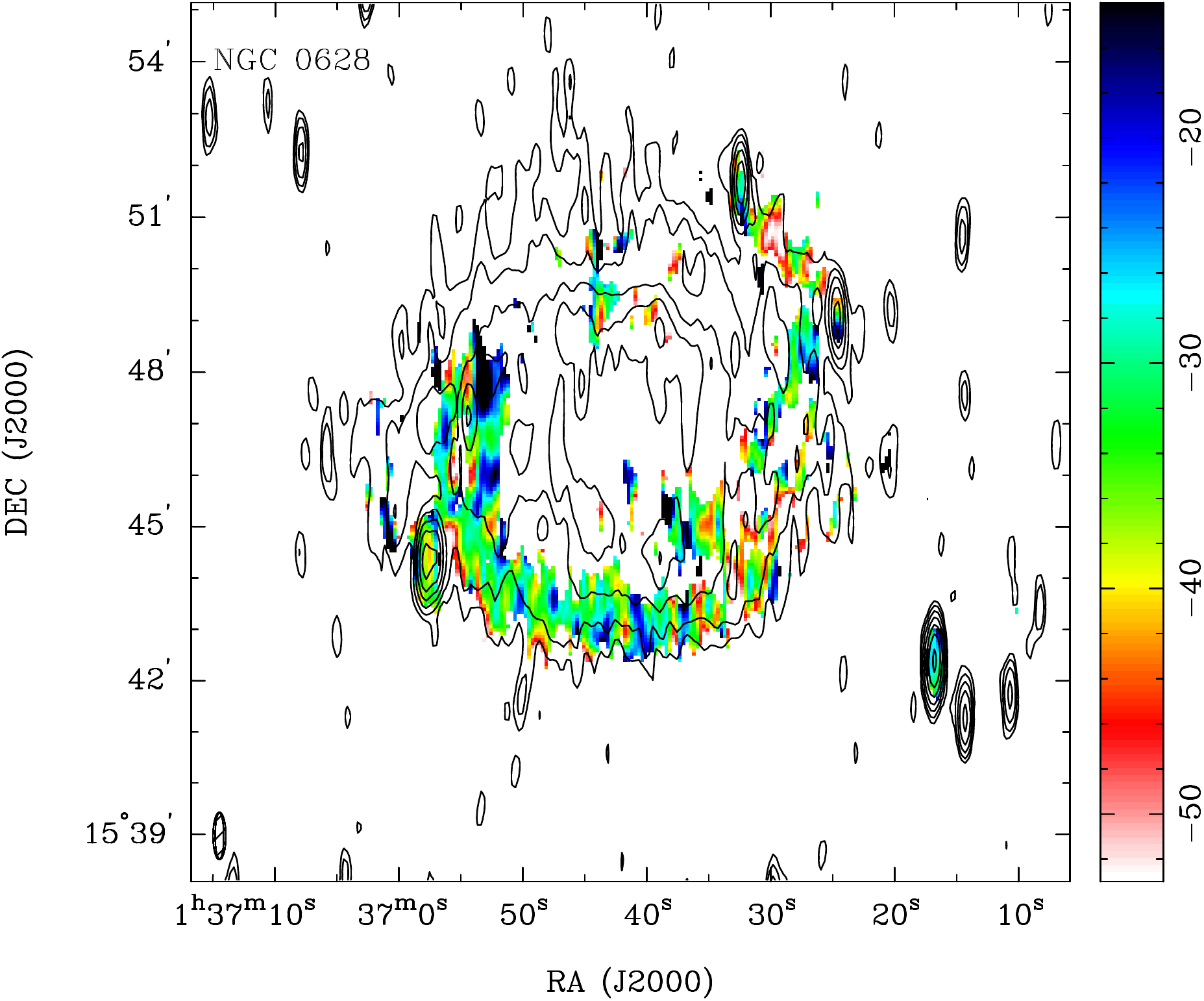}\includegraphics{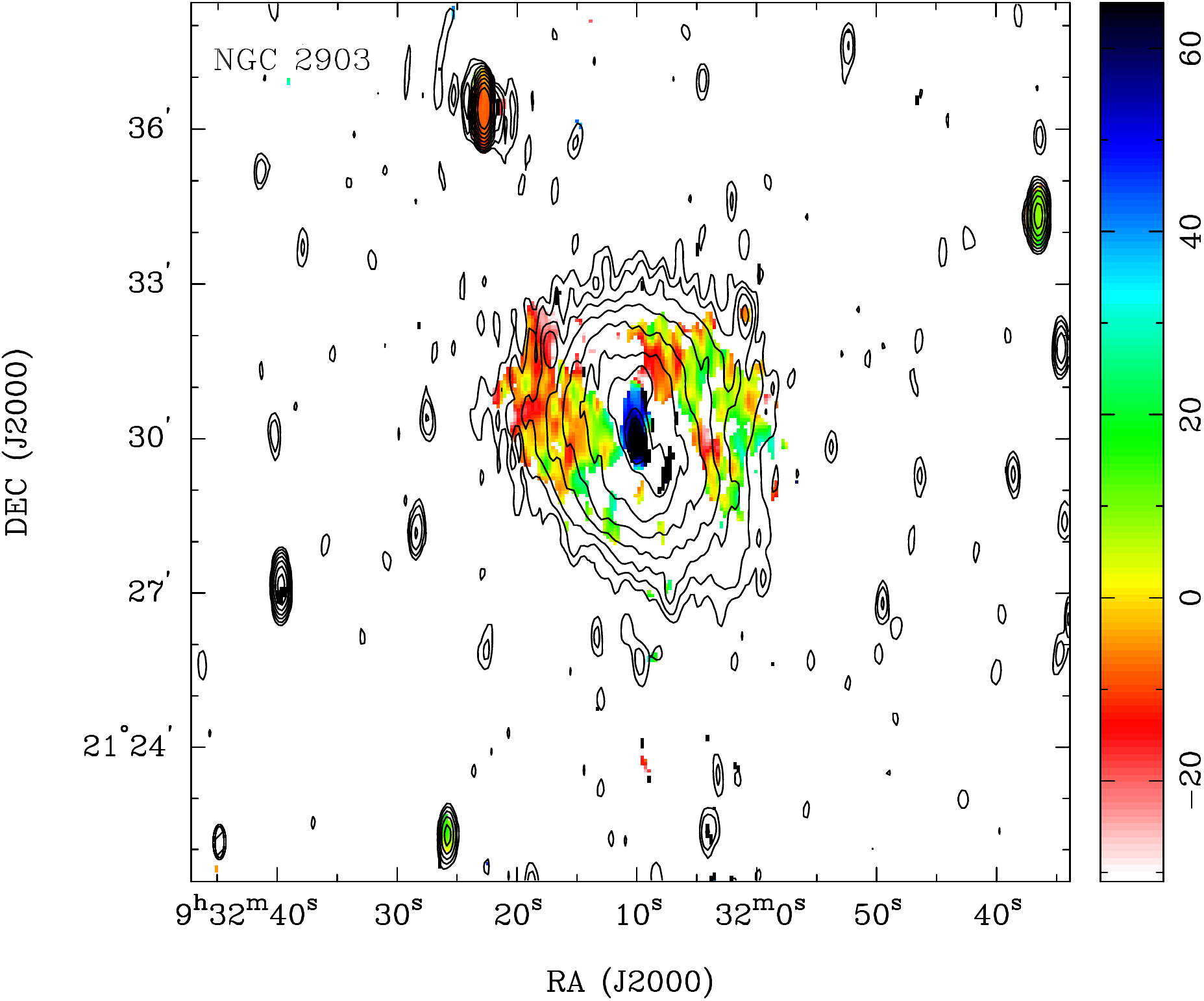}}
\resizebox{\hsize}{!}{\includegraphics{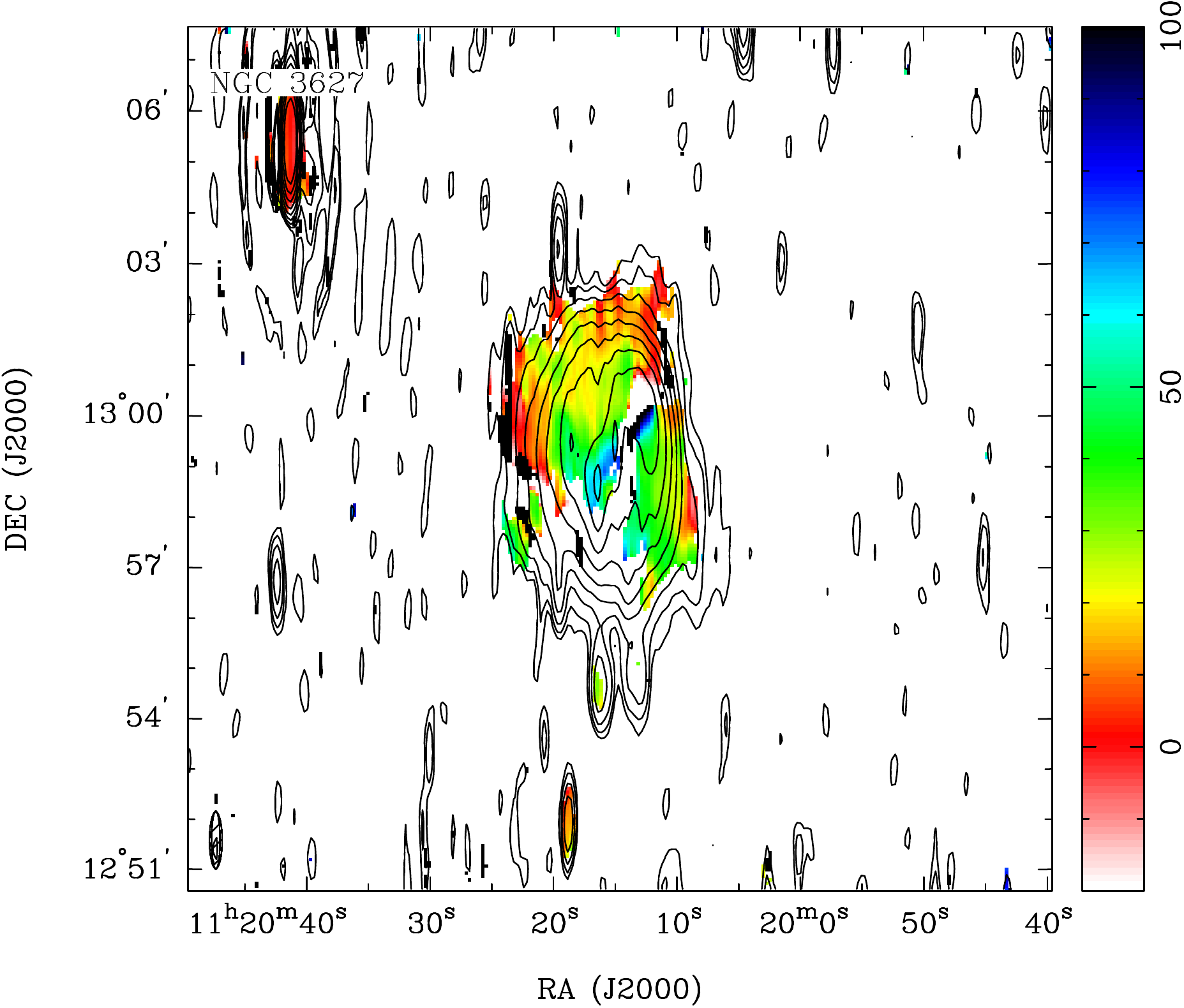}\includegraphics{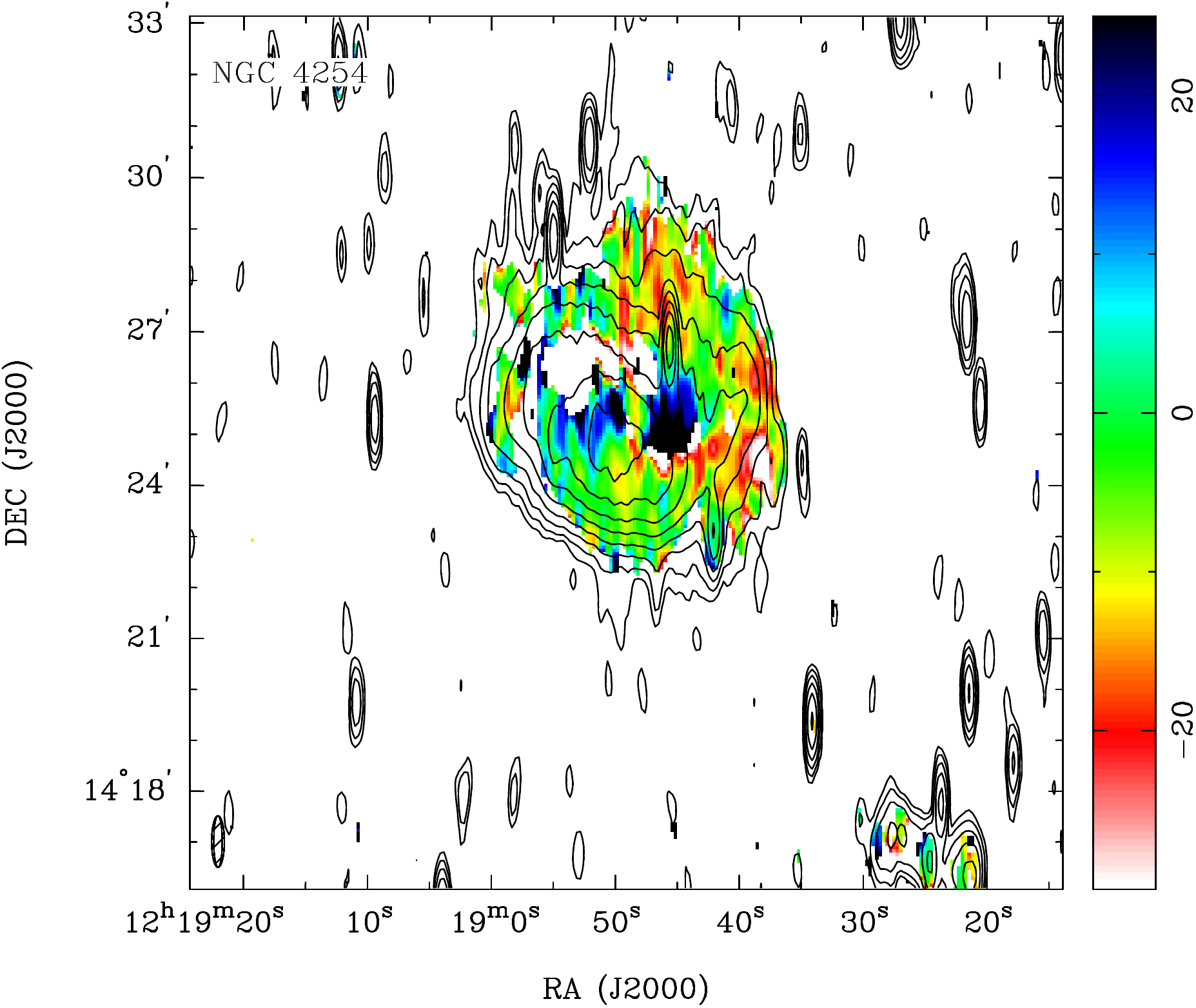}}
\resizebox{\hsize}{!}{\includegraphics{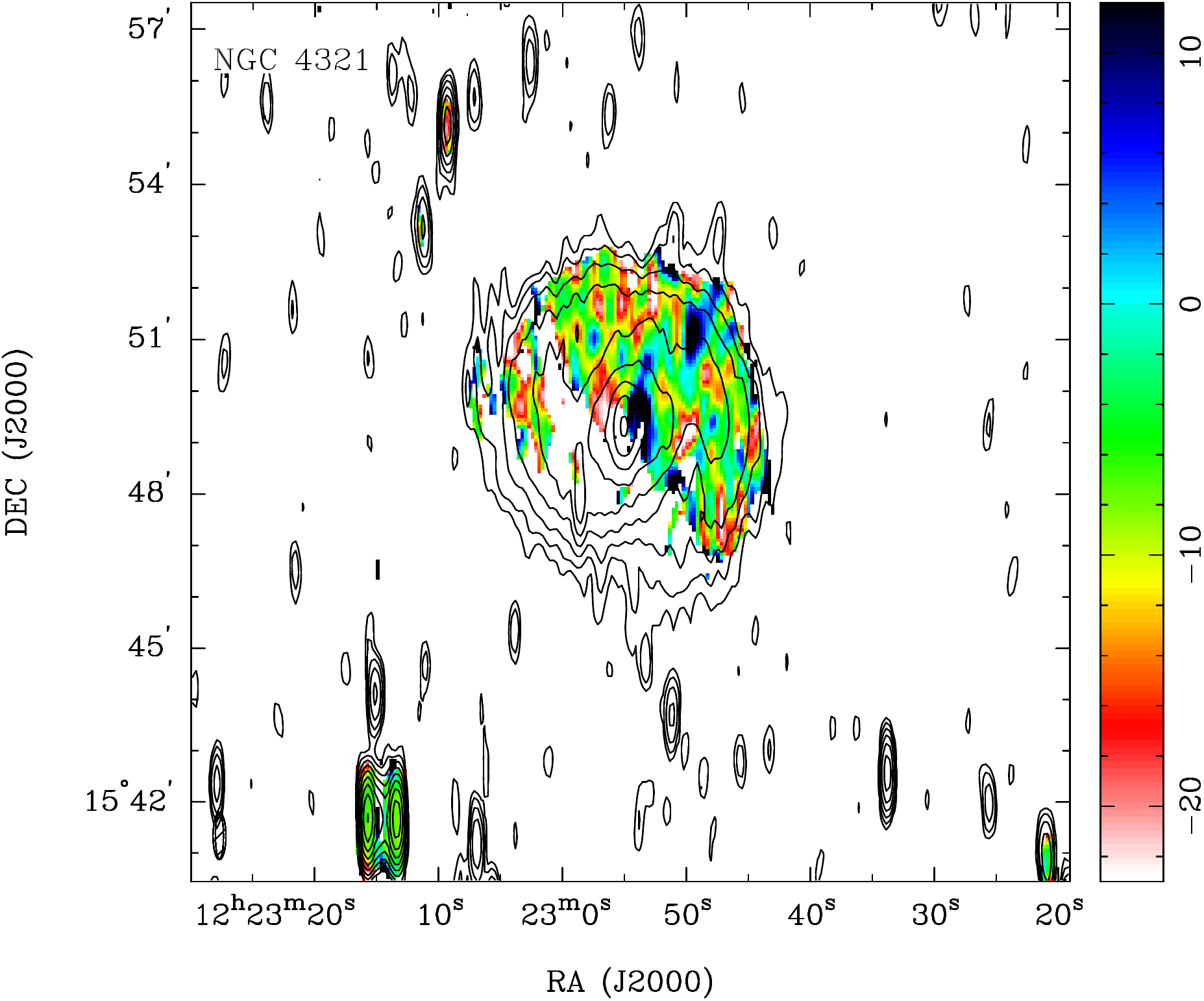}\includegraphics{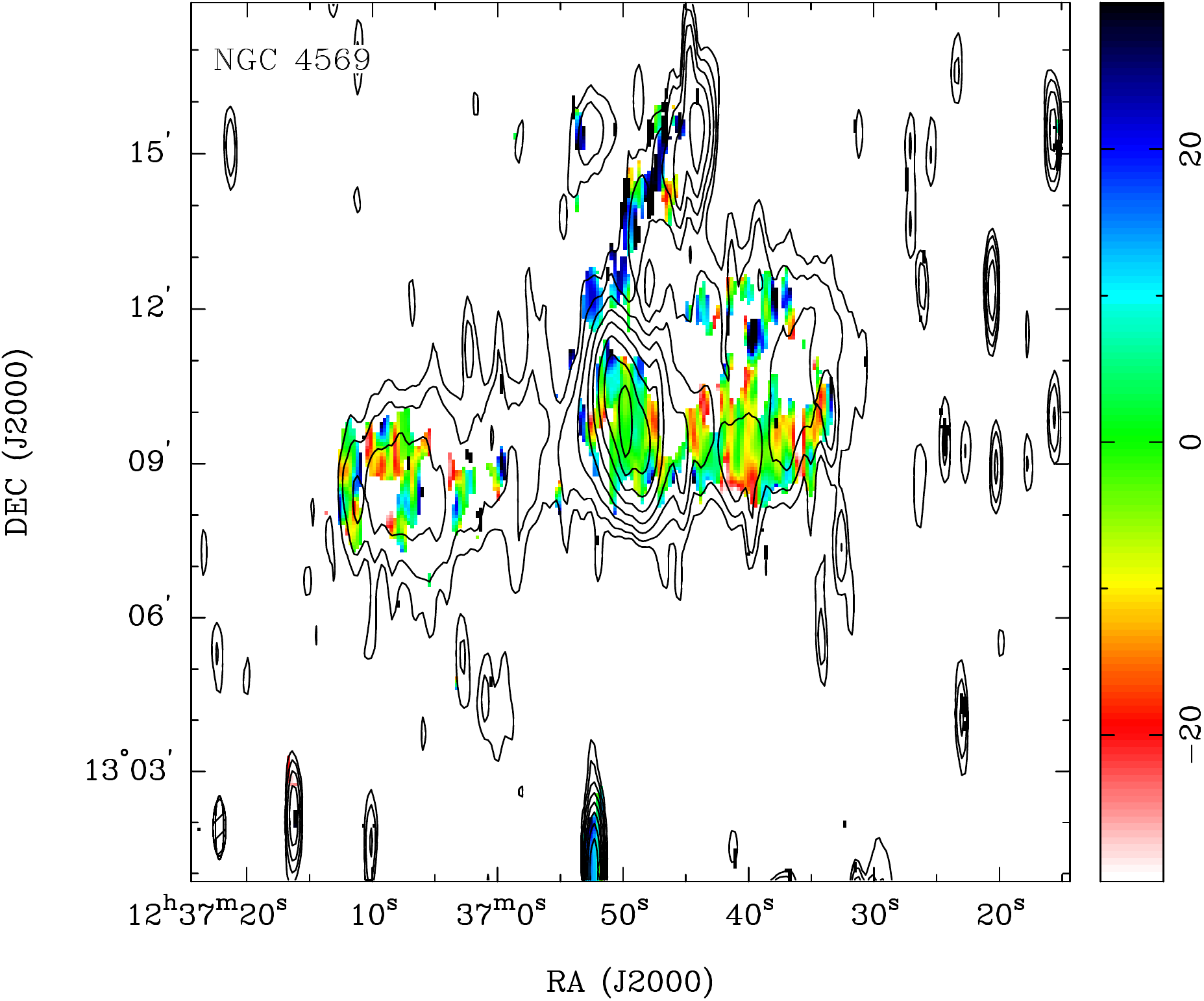}}
\caption{Images of peak $\phi$ in some of the survey galaxies. The
  colorbar at the top edge shows the range of $\phi$ [rad/m2]
  displayed with the colormap. Galaxy ID is indicated in the upper left of each panel.}
\label{figure:pkphi}
\end{figure*}

\addtocounter{figure}{-1}
\begin{figure*}
\resizebox{\hsize}{!}{\includegraphics{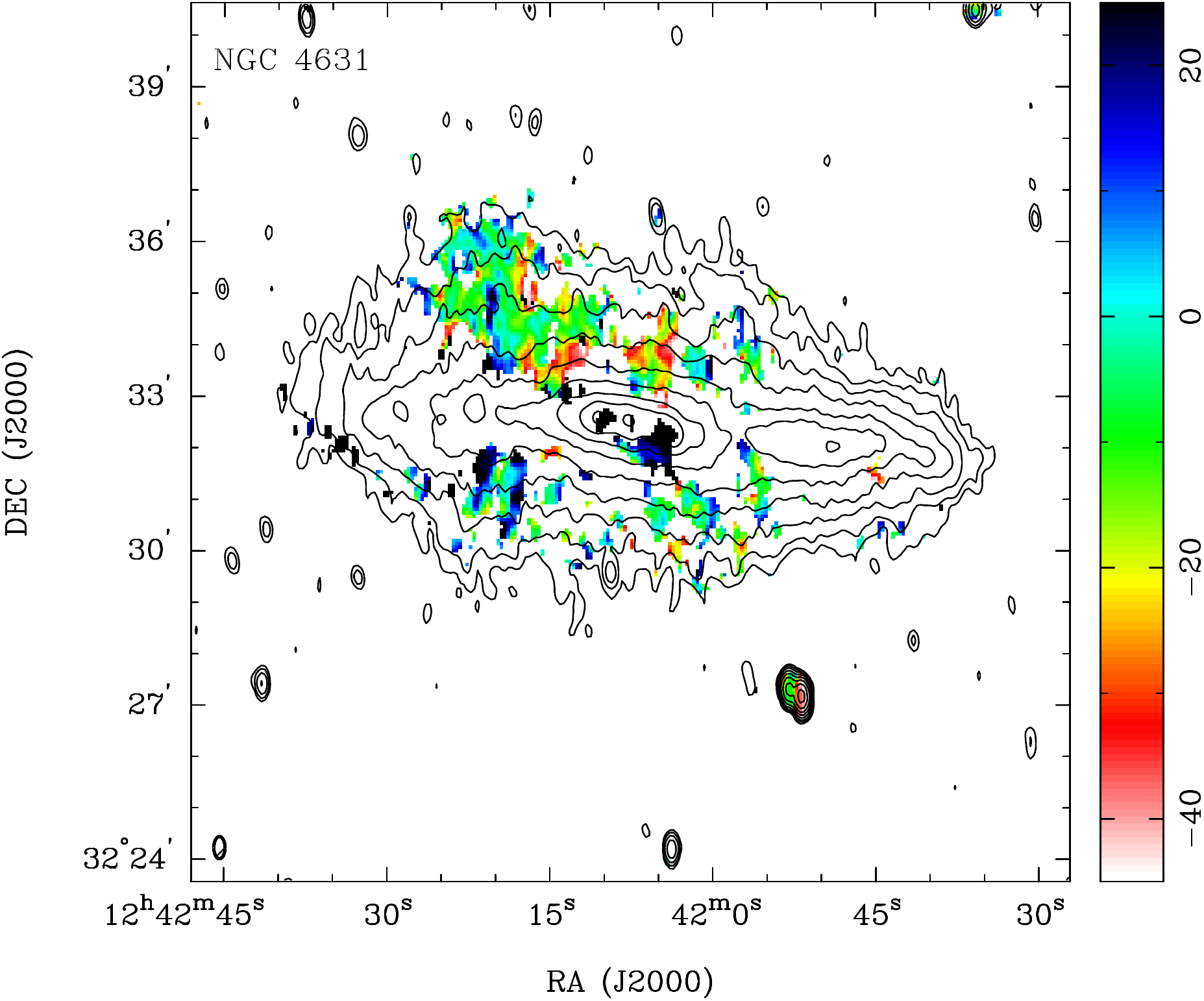}\includegraphics{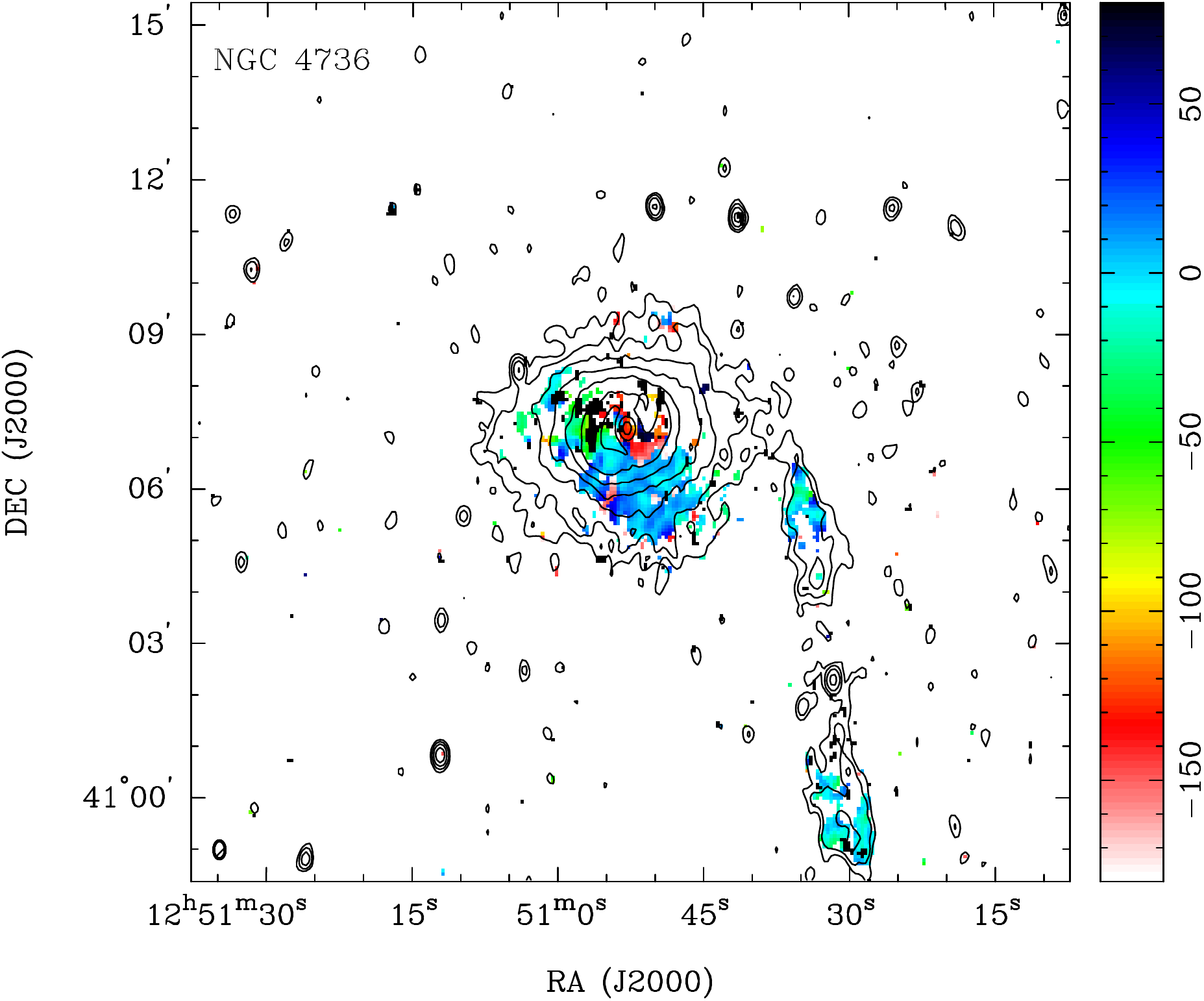}}
\resizebox{\hsize}{!}{\includegraphics{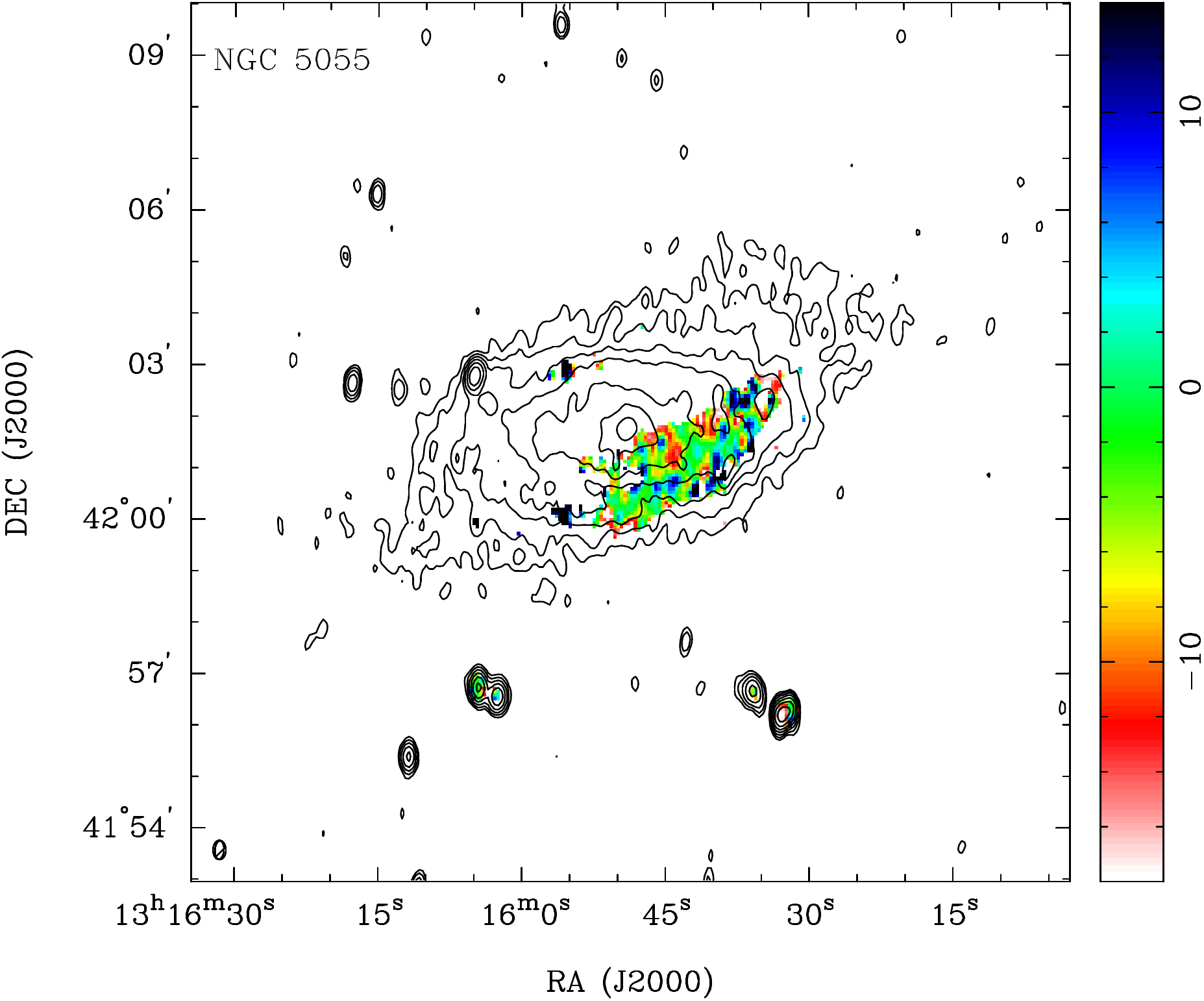}\includegraphics{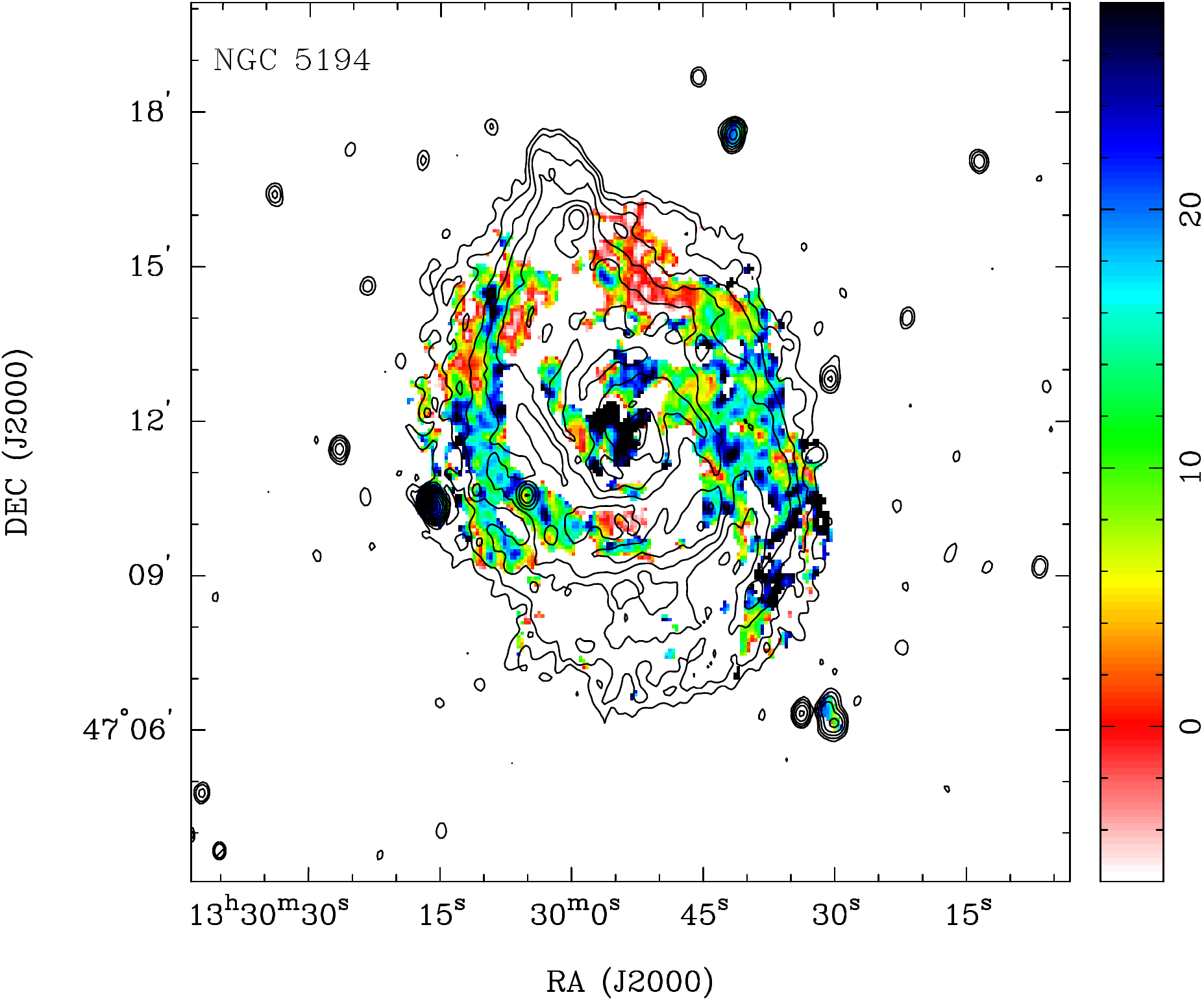}}
\resizebox{\hsize}{!}{\includegraphics{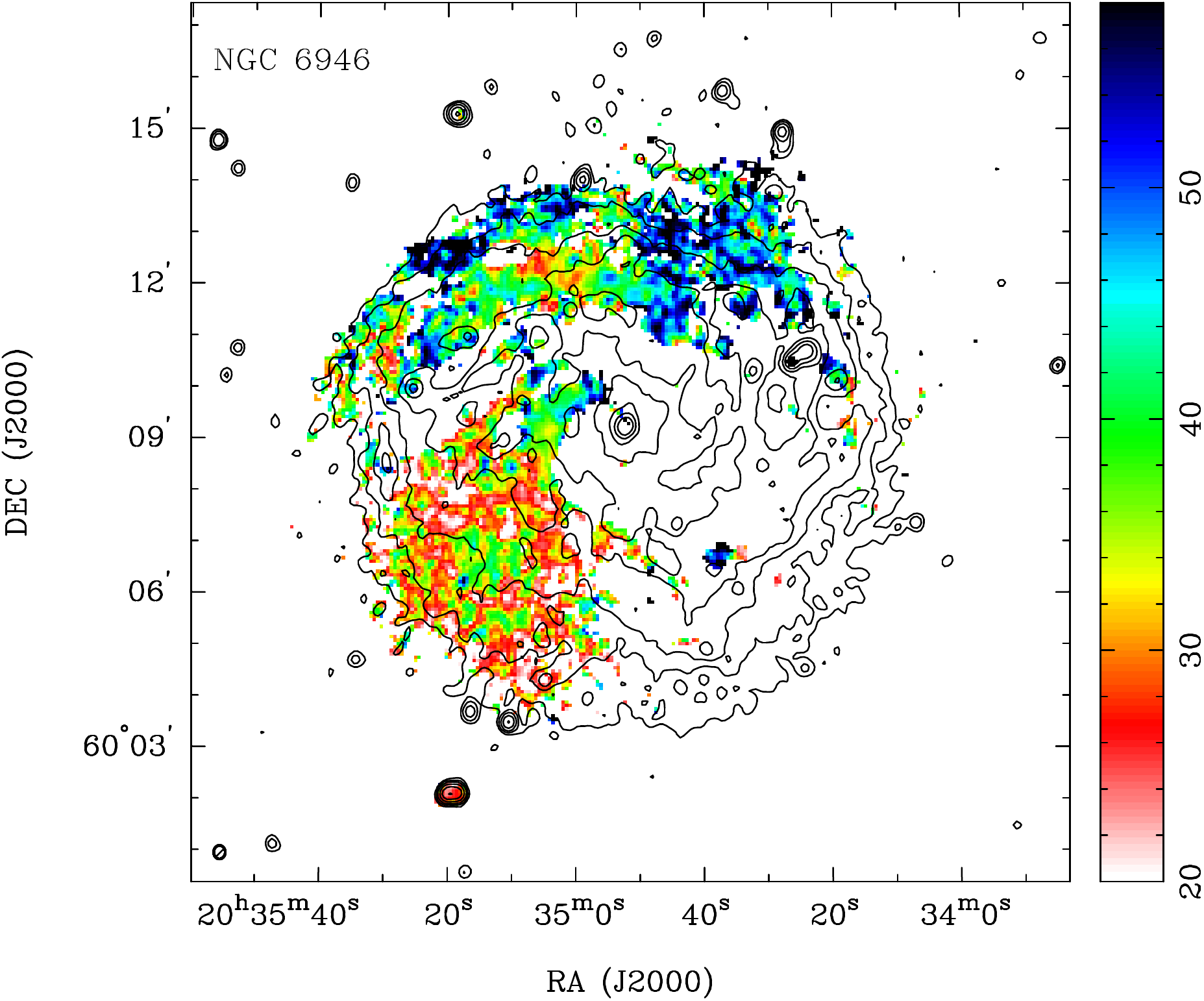}\includegraphics{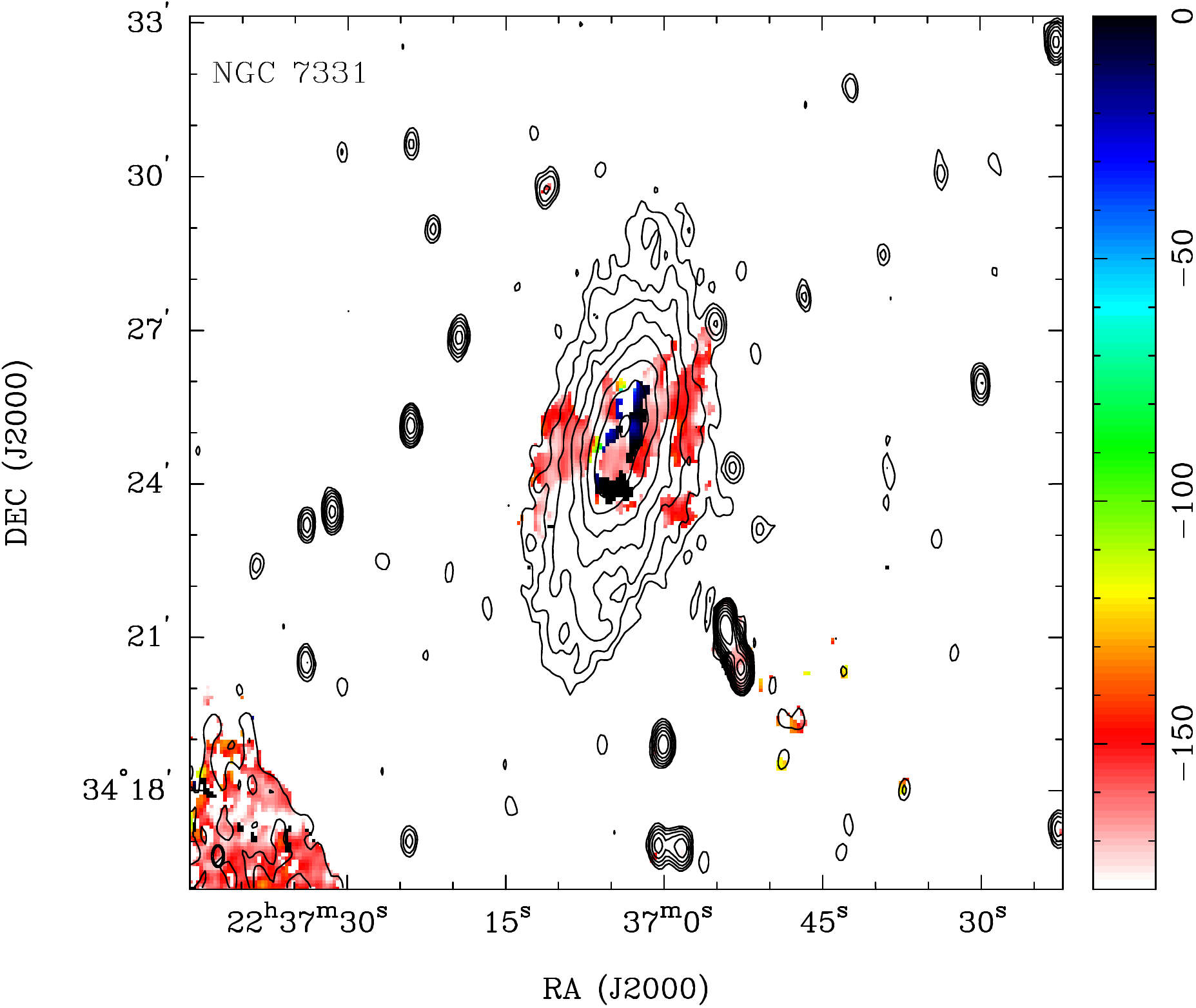}}
\caption{(continued) Images of peak $\phi$ in some of the survey galaxies.}
\end{figure*}

\section{Discussion\label{section:discussion}}

\subsection{Circular polarization}

In cases of extreme Faraday rotation, linearly polarized emission can
be {\em Faraday converted} into circular polarization
\citep[e.g.,][]{jones_odell_1977}. Stokes $V$ images (created as part
of the pipeline described in \S\,\ref{subsection:standardreductions})
were examined for signs of any circularly polarized emission
associated with the target galaxies. No detections were made.

\subsection{Magnetic field distributions\label{subsection:Bdist}}

In \S\,\ref{subsection:notes}, a few general trends can be discerned
in the sample spiral galaxies. The most obvious of these is that the
polarized intensity is minimized along the receding major axis. This
points to a common global magnetic field geometry which is tied not
only to the morphology of the galaxy, but also to the dynamics of the
galaxy. In Paper III, we discuss a quite general and simple model
which may be at the origin of the observed patterns.

\subsection{Extended Faraday depth profiles\label{subsection:extprof}}

The frequency coverage obtained in the WSRT-SINGS survey is sufficient
for excellent recovery of polarized emission at a single Faraday depth
or multiple well-separated Faraday depths. Well-separated regions
of synchrotron emission and Faraday rotation lead to ``Faraday thin''
emission, which appears as one or more
unresolved features in the Faraday dispersion function. The WSRT-SINGS
frequency coverage is however insufficient for
recovery of polarized emission at a continuous range of
Faraday depth. Such circumstances occur in regions referred
to as ``Faraday thick''. In Faraday thick regions, emitting and
rotating plasmas with regular magnetic fields may be uniformly
collocated along the LOS, such that synchrotron emission from the far
side of the volume suffers more Faraday rotation than the synchrotron
emission from the near side.
In the simplest such case, a constant level of polarized flux will be
detected at a continuous range of Faraday depth. Faraday thickness can
also originate in volumes in which the Faraday rotation is generated
by turbulent magnetic fields. This will also lead to polarized
flux being distributed over a range of Faraday depth \citep{burn_1966}. 
\citet{berkhuijsen_etal_1997} describes how this latter mechanism
can cause depolarization of the synchrotron emission within the disk at
18- and 20-cm wavelengths in the particular case of NGC 5194. In this
picture, the intervening halo is transparent to polarized emission,
and acts as a ``Faraday screen''.

Recall that the resolution in Faraday depth space, $\Delta\phi$, is
inversely proportional to the width of the sampling in $\lambda^2$
space, $\Delta\lambda^2$. The offset of the $\lambda^2$ sampling from
$\lambda^2=0$ does not affect the resolution. However, the ability to
detect polarization in the presence of internal Faraday depolarization
(either caused by regular or turbulent fields, as described above)
is determined by the actual values of $\lambda^2$. As shown by
\citet{burn_1966}, when observing a Faraday thick region with regular
magnetic fields, the $||P(\lambda^2)||$ distribution is a sinc function.
Differential Faraday rotation within the emitting and
rotating region depolarizes the emission to some degree at all
non-zero wavelengths, and the effect is generally stronger at larger
$\lambda^2$.
The depolarization takes place within the volume and
not at the telescope. It is thus independent of the channel
width used in performing the observation, but it is dependent on the
frequency band itself.
The fractional recovery of the polarized flux
by RM-Synthesis is determined by the
sampled values of $\lambda^2$. A smaller value of $\lambda_{\mathrm{min}}^2$
means sampling $||P(\lambda^2)||$ closer to its peak, and thus recovering
more of the intrinsic polarized flux.
With observations made at $\lambda^2\gg0$,
neither the RM-Synthesis technique nor a subsequent {\tt RM-CLEAN}
operation will completely recover the intrinsic degree of polarization
(i.e., $P(\lambda^2=0)$) if internal depolarization has been
present. There is no substitute for obtaining the required full
sampling of $\lambda^2$.

The reconstruction of the intrinsic polarized flux in Faraday thick
regions is incomplete because
of the Fourier transform at the heart of RM-Synthesis. Large-scale structures
in $P(\phi)$ are recovered by observations at small $\lambda^2$. Thus the
consequence of $\lambda_{\mathrm{min}}^2\gg0$ in a given observation is
that only the high-frequency structures in $P(\phi)$ are sampled. The
front and back ``skins'' of a Faraday thick region will be detected in
polarization, each with a depth in $\phi$ of about
$\pi/\lambda_{\mathrm{min}}^2$ (the largest scale that we are able
to recover with our frequency sampling). For the current observations with
$\lambda_{\mathrm{min}}$~=~17~cm, this corresponds to a skin depth of
about 108~rad~m$^{-2}$, while our resolution is about
$\Delta\phi\,\approx\,\rpms{144}$. If we imagine a uniform slab with a
Faraday depth exceeding $2\pi/\lambda_{\mathrm{min}}^2$, we might
begin to resolve the front and back skins of such a structure, albeit
with the inevitable reduction of polarized intensity from its
intrinsic value.

Although we are unable to reconstruct the intrinsic degree of
polarization for arbitrary Faraday thick structures with the present
frequency coverage, we can look for indications that Faraday thick
regions are present. The most obvious of these would be the detection
of $\phi$-broadening, or even resolved $\phi$-splitting in Faraday
depth. The amount of broadening or splitting would begin to constrain
the likely degree of depolarization that affects the current
observations. A systematic decrease of polarized emission, such as
seen toward the southwest half of NGC 6946, would be a more ambiguous
indicator. This form of differential depolarization would require a
systematic increase in the Faraday depth of some regions relative to
others. In the case of NGC 6946 this seems rather unlikely to be due
to the distribution of electron density, as also concluded by
\citet{beck_2007}, but may instead be due to a large-scale pattern in
the field geometry which might lead to both a systematic increase of
the Faraday depth as well as a decrease in the intrinsic degree of
polarization (given their orthogonal dependence on the field
orientation). We return to this discussion in Paper III.

To assess the presence of broadened and/or split Faraday dispersion
function profiles, we adapt the so-called ``velocity coherence''
technique described by, e.g., \citet{braun_etal_2009}. The {\tt
  RM-CLEAN}ed $P(\phi)$ cubes were smoothed along the $\phi$ axis with
a boxcar kernel, of width $\rpms{143}$, which is similar to the FWHM
of the Faraday resolution. After this smoothing operation,
an unresolved Faraday dispersion function will peak at about $81\%$ of
its original amplitude. Broadened profiles will have a higher relative
peak amplitude. Images of the Faraday depth coherence,
$\phi_{\mathrm{C}}\,\equiv\,P(\hat{\phi}_{\mathrm{bc}})/P(\hat{\phi})$,
were produced for each galaxy, where $P(\hat{\phi}_{\mathrm{bc}})$ is
the peak polarized flux in each pixel of the boxcar-smoothed $P$ cube,
and $P(\hat{\phi})$ is the same quantity in the original
cube. Inspection of these images did not reveal any global systematic
patterns, but some small-scale localized features are of note.

In galaxies with distributed polarized flux, small localized Faraday
thick regions tend to appear in interarm regions. In NGC 4254, the
region between the nucleus and the large northwestern spiral arm is
significantly Faraday thick compared to the rest of the disk. In NGC
4631, the average $\phi_{\mathrm{C}}$ is somewhat higher than in the
other sample galaxies. This is not unexpected, since its edge-on
orientation may cause a significant amount of Faraday depolarization. The
largest values of $\phi_{\mathrm{C}}$ (corresponding to the broadest
$P(\phi)$ profiles) appear along the southern edge of the polarized
extension in the northeast quadrant of the galaxy. In other targets,
there does not appear to be a recognizable structure in
$\phi_{\mathrm{C}}$ which can be associated with morphological
features.

\subsection{Nuclear Faraday dispersion functions\label{subsection:nuc}}

\begin{figure}
\resizebox{\hsize}{!}{\includegraphics{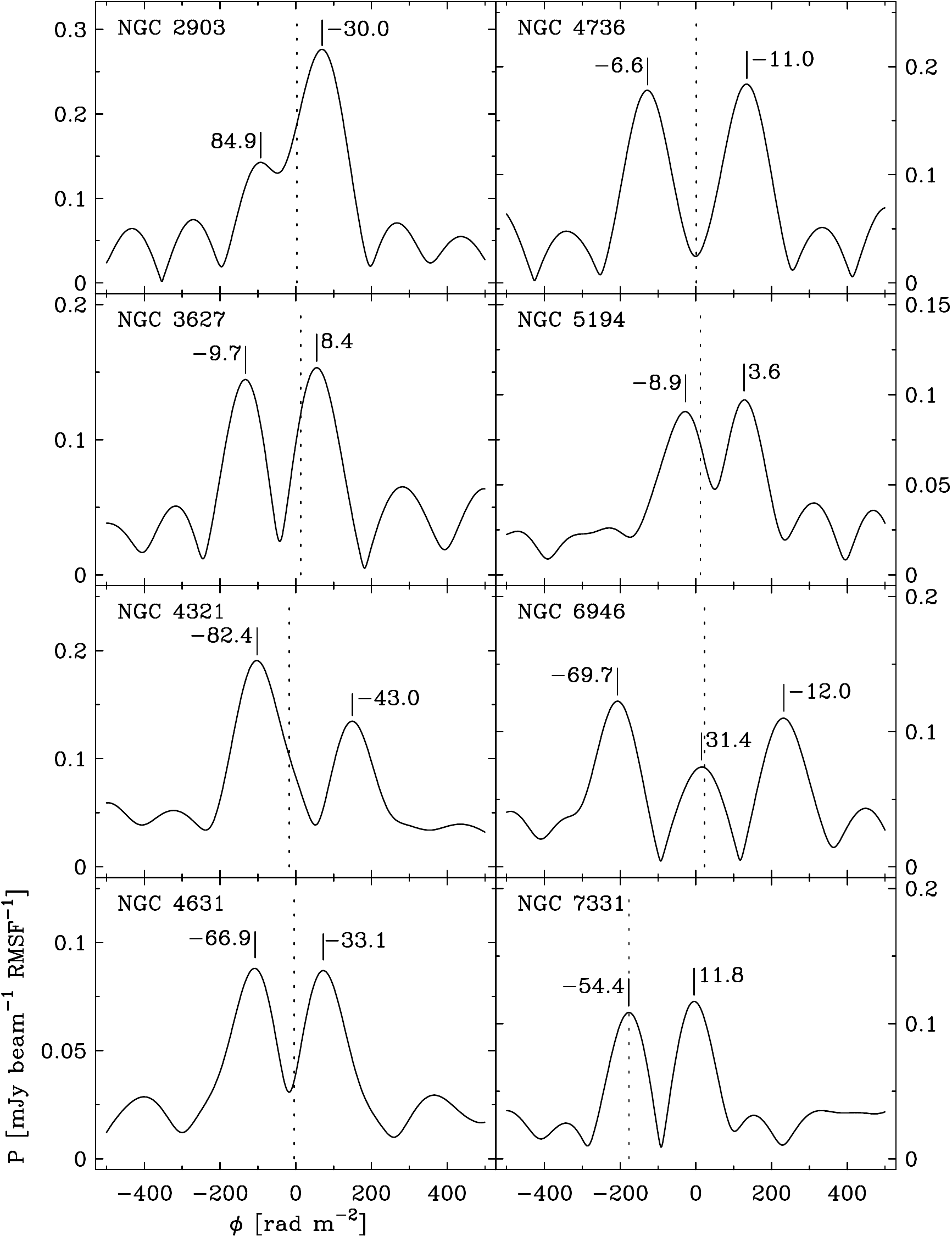}}
\caption{RM spectra extracted from the central regions of the
  indicated galaxies (coordinates are listed in Table
  \ref{table:centspec}). Features with tabulated characteristics are
  indicated with solid lines; the polarization angle at that value of RM is
  indicated. The Galactic foreground contribution (from Table
  \ref{table:summary}) is indicated with dotted lines.}
\label{figure:centsp}
\end{figure}

In all of the galaxies with compact, (circum-)nuclear polarized emission,
the nucleus is detected at two distinct Faraday depths (and perhaps three in the
case of NGC 6946) which are offset to both positive and negative
values from the estimated Galactic foreground RM by about
$\rpms{100}$. Recall that this splitting corresponds almost exactly
with the condition noted above for resolving the two skins of an
intrinsically Faraday deep distribution with our observing setup, of
$2\pi/\lambda_{\mathrm{min}}^2~\sim~\rpms{200}$. A possible exception
is NGC 4569, for which the polarized emission appears to be associated with the
nucleus, but shows only a hint of a broadened feature in the Faraday
dispersion function. Table \ref{table:centspec} lists the
characteristics of the polarized nuclear emission in the targets with
polarized emission at multiple Faraday depths. The columns are
(1) Galaxy ID; (2,3) RA (J2000.0) and Dec. (J2000.0) from which
the Faraday dispersion functions were extracted; (4) Nucleus
classification from \citet{ho_etal_1997}, where H indicates HII
nucleus, S indicates Seyfert nucleus, L indicates LINER nucleus,
T indicates transition nucleus, and numbers indicate subclasses; (5) RM
feature number; (6) $\phi$ value at the location of the RM feature;
(7) Total polarized flux at that RM; (8,9) Stokes Q and U values
at that RM; (10) Polarization angle derived from Q and U.
The deconvolved Faraday
dispersion functions are plotted in Figure \ref{figure:centsp}. Note
that in Figure \ref{figure:images:a}, only the polarized flux (and
polarization angle) of the brightest of these components is plotted.

\begin{table*}
\caption{Characteristics of Split RM Profiles in Galaxy Cores.}
\label{table:centspec}
\centering
\begin{tabular}{l l l c c c c c c c }
\hline\hline
Galaxy & RA & Dec & Nucleus & Feature & $\phi$ & $P(\phi)$ & $Q(\phi)$ & $U(\phi)$ & $\chi(\phi)$ \\ 
ID & J2000 & J2000 & Type & no. & [$\rpms{}$] & [$\mu$Jy] & [$\mu$Jy] & [$\mu$Jy] & [deg.] \\
\hline
NGC 2903 & 09 32 09.8 & 21 30 00 & H     & 1 &  -93 & 143.0 &  -102.4 &   18.3 &  84.9 \\
NGC 2903 & $\,$       & $\,$     & $\,$  & 2 &  +70 & 276.0 &   141.3 & -245.6 & -30.0 \\
NGC 3627 & 11 20 15.8 & 12 59 26 & T2/S2 & 1 & -133 & 144.0 &   111.7 &  -39.5 &  -9.7 \\
NGC 3627 & $\,$       & $\,$     & $\,$  & 2 &  +55 & 153.0 &   146.8 &   44.1 &   8.4 \\
NGC 4321 & 12 22 54.9 & 15 49 21 & T2    & 1 & -103 & 191.0 &  -149.6 &  -40.6 & -82.4 \\
NGC 4321 & $\,$       & $\,$     & $\,$  & 2 & +150 & 135.0 &     9.6 & -136.6 & -43.0 \\
NGC 4631 & 12 42 07.7 & 32 32 26 & H     & 1 & -108 &  88.0 &   -57.5 &  -60.1 & -66.9 \\
NGC 4631 & $\,$       & $\,$     & $\,$  & 2 &  +73 &  87.0 &    33.1 &  -74.9 & -33.1 \\
NGC 4736 & 12 50 53.1 & 41 07 14 & L2    & 1 & -128 & 178.0 &   268.3 &  -63.1 &  -6.6 \\
NGC 4736 & $\,$       & $\,$     & $\,$  & 2 & +135 & 184.0 &   164.6 &  -66.5 & -11.0 \\
NGC 5194 & 13 29 52.7 & 47 11 47 & S2    & 1 &  -28 &  91.0 &    12.7 &   -4.1 &  -8.9 \\
NGC 5194 & $\,$       & $\,$     & $\,$  & 2 & +128 &  97.0 &    73.7 &    9.3 &   3.6 \\
NGC 6946 & 20 34 52.3 & 60 09 22 & H     & 1 & -208 & 122.0 &   -53.1 &  -45.4 & -69.7 \\
NGC 6946 & $\,$       & $\,$     & $\,$  & 2 &  +15 &  74.0 &    26.2 &   50.9 &  31.4 \\
NGC 6946 & $\,$       & $\,$     & $\,$  & 3 & +233 & 110.0 &    68.3 &  -30.4 & -12.0 \\
\hline
\end{tabular}
\end{table*}

What is the origin of the split Faraday dispersion functions? Given
the $\lambda^2$ coverage of the observations, there are two
possibilities. Either polarized emission with two distinct rotation
measures originates within the spatial beam of our observations, or we
are detecting Faraday thick emission. In the latter case (as noted in
\S\,\ref{subsection:extprof}), our observational setup is unable to
recover extended features in the Faraday depth domain. We would
recover emission only from the edges of such a structure, giving the
appearance of two distinct features. In either case, the physical
origin must be a characteristic geometry of the nuclear magnetic field and
ionized gas distribution common to all galactic nuclei with such
Faraday dispersion functions. Understanding these nuclear Faraday
dispersion functions may therefore illuminate the physics in the
central regions of these galaxies. The rather general detection of
both positive and negative net Faraday depths is already indicative
of both positive and negative signs of $B_\parallel$. Such a sign
change of the LOS field, seen from galaxies inclined between 30 and 85
degrees, might be understood with a radially directed (outward or inward) field
geometry in either or both of the galaxy mid-plane or parallel to the
rotation axis. We note that somewhat higher resolution observations of
NGC 6946 \citep[][see his Figure 11]{beck_2007} show a continuation of
the spiral pattern all the way into the nuclear regions, but perhaps with
a greater radial component than at larger radii. Since, in the nuclear
Faraday dispersion functions, comparable levels of polarized intensity
are seen from both the negative and positive RM components, it seems
plausible that the synchrotron emitting region has a significant radial
extent that is relatively symmetric about the origin of the sign change of
$B_\parallel$, which presumably corresponds to the galaxy nucleus.
It is worth mentioning that a global magnetic field geometry which is
everywhere inward- or outward-directed would, in the spatially unresolved
central region, show at least a double-peaked Faraday dispersion function.
However, the magnitude of the $\phi$ values that we observe in the nuclear
regions are much larger than the typical $\phi$ values elsewhere in the
disks, suggesting either that an additional magnetic field component
is present in the nuclei, or that enhanced magnetic field strengths
and/or electron densities are present in the nuclear regions.

In the case of NGC 4631, the split Faraday dispersion function may be
influenced by the almost edge-on aspect of the galaxy. The orientation
could lead to a large Faraday depth toward the nucleus, which might
also be responsible for the prominent areas of likely depolarization
throughout the disk plane. A connection of Faraday depth with the
outer disk orientation is not as obvious for the more face-on
targets. In fact, the largest observed splitting between the two
detected components is seen in NGC~6946, for which the disk
inclination is only 33$^\circ$. For a circum-nuclear dipole field, a
sign reversal would not occur in the vertical component, but
would only be expected to occur in the radial
component. This would be most apparent in an edge-on viewing
geometry. For a quadrupole field a sign reversal would be
expected to occur in both the radial and vertical directions,
and thus might be seen independent of the viewing angle. Although
neither edge-on nor face-on viewing geometries of the outer disk are
strongly preferred for witnessing this phenomenon, our sample size is
not large enough to draw general conclusions on a preferred
circum-nuclear field configuration. There is also no immediately
apparent trend with the type of nucleus. Nuclei with optical
\citep{ho_etal_1997} and MIR \citep{dale_etal_2006} emission
consistent with an AGN are just as likely to show split
profiles (4/11) when compared to nuclei which are classified as
non-AGN (3/10). Similar analysis, using
the RM-Synthesis technique, of a larger number of galaxies would
help to clarify the situation.

Based on our sample, a sufficient condition for the existence of a
multiply peaked nuclear Faraday disperion function is simply the
presence of a compact, polarized nuclear source. There are two
apparent exceptions to this pattern which deserve further
comment. NGC~7331 has split profiles but no compact nuclear
source. This particular galaxy exhibits split Faraday dispersion
functions throughout the inner disk, so the splitting noted here does
not apply to a nucleus. Moreover, only one of the RM components in
NGC~7331 (the inner disk) is displaced significantly from the Galactic
foreground value, rather than both as seen toward polarized nuclear
sources. On the other hand, NGC 4569 has what appears to be a compact,
polarized nuclear source, but the nuclear Faraday dispersion function
is not obviously split. In this case there is some indication for
broadening of the profile toward the nucleus in the Faraday depth
coherence image described in \S\,\ref{subsection:extprof};
it may be that there is splitting which is
only barely resolved with our RM resolution ($\rpms{\sim144}$). 
Further work to investigate nuclear polarization should be
performed at higher frequency, where the effects of internal
Faraday depolarization are less important, and Faraday thick regions
can be unambiguously distinguished from regions with multiple distinct
Faraday depths. At higher frequencies, the width of the RMSF will
likely be broader than we have obtained here. Judicious combinations of
observations at low and high frequency can lead to the recovery of
Faraday thick polarized flux. As pointed out by
\citet{brentjens_debruyn_2005}, the condition for recovering such flux
is $\lambda_{\mathrm{min}}^2<\Delta\lambda^2$.

\section{Conclusions and Outlook\label{section:conclusions}}

We have presented linear polarization data measured in two broad
frequency bands near 1400 and 1700 MHz in the WSRT-SINGS survey. The
RM-Synthesis method was used to reconstruct the intrinsic properties
of the polarized emission, obtaining and correcting for the Faraday
depth contributions from both the Milky Way and the target galaxies
themselves. The reconstructed Faraday dispersion functions were
deconvolved using a technique similar to the {\tt CLEAN} algorithm
commonly utilized in synthesis imaging. The deconvolution was
particularly important with the present observations, because
the gap in frequency coverage between our two frequency bands causes the
RM spread function (RMSF) to exhibit sidelobes at nearly the 80\%
level.

The results of these processing steps were used to derive maps of the
linearly polarized flux in each of the target galaxies, and, in cases
where sufficient flux was measured, to analyze the spatial
distributions of Faraday rotation measures, which probe the component
of the magnetic field along the line of sight, and polarization
vectors, which probe the component of the magnetic field perpendicular
to the line of sight. The Faraday rotation corrected polarization
angles were used to generate maps of the magnetic fields perpendicular
to the line of sight.

Linearly polarized emission was detected in 21 of the 28 galaxies
considered in this investigation. The detected galaxies all have
Hubble type between Sab and Sd; only three galaxies (out of 24) in
this range of classification were undetected in polarized flux.  All
of the (albeit few) sample galaxies with Hubble type later than Sd or
earlier than Sa were all undetected. We have not detected any
circularly polarized emission from any of the galaxies.

The most prominent trend which has emerged through this analysis is
that in all galaxies with spatially extended
polarized emission, the azimuthally-binned polarized flux is
consistently lowest along the receding major axis. For such a trend to
appear in such a large and diverse sample of spiral galaxies implies
that a common magnetic field geometry has been revealed. In Paper III,
we attempt to model the observed azimuthal variations in both rotation
measure and polarized flux, using toy models of axi- and bi-symmetric
magnetic field configurations, with the additional possibility of a
non-zero vertical component to the field. We find that such a magnetic
field geometry can explain the azimuthal variation in polarized flux,
its dependence on inclination, as well as the azimuthal variation in
the rotation measures attributed to the target galaxies.

Another interesting feature was discovered in the galaxies with
prominent nuclear emission in both total power and linearly polarized
flux. The Faraday dispersion functions in those galaxies' cores show
indications of significant broadening and/or splitting, indicating the
presence either of spatially collocated synchrotron-emitting and
Faraday-rotating plasma, or distinct Faraday thin emitting regions
within the resolution element. At the wavelengths observed in this
survey, we are rather insensitive to significantly broadened Faraday
dispersion functions -- these would appear (at best) as double-peaked
profiles. It is possible that all of these galaxies host {\it Faraday
  thick} regions in their cores, but we are unable to make this
distinction. Observations at higher radio frequencies, where
depolarization issues are less significant, would clarify the
situation. What is clear, from the occurence of both positive and
negative net RMs (after accounting for the Galactic foreground
contribution), is that the polarized emission arises on either side of
a reversal of the LOS field, presumably reflecting a radially directed
field (either inward or outward) centered on the nucleus. Whether
the LOS magnetic field simply changes sign near the nucleus, or
polarized flux is present at an extended range of Faraday depth
(centered near $\phi=\rpms{0}$), is not clear with the present
observations.

The techniques used here can be extended to observations performed at
other radio telescopes. In particular, the new class of telescopes
which are being built now and into the era of the SKA, will all
provide polarization data at excellent frequency resolution covering
wide bandwidth. These telescopes will provide data which is
excellently suited for use with the RM-Synthesis method, and will
enable the study of a larger sample of galaxies at greater
sensitivity, and improved resolution in the Faraday domain. Of
particular interest is the LOw Frequency ARray
\citep[LOFAR;][]{falcke_etal_2007} array which is presently being
built, and will operate at frequency ranges between $30-80$ and
$120-240$ MHz. At these low frequencies, we will have RM resolutions
of order $\rpms{1}$ and better (cf. the WSRT-SINGS RM resolution
$\rpms{\approx144}$), yielding sensitivity to far weaker magnetic field
structures in nearby galaxies \citep{beck_2008}.

\begin{acknowledgements}
We thank Michiel Brentjens for providing his RM-Synthesis software,
and for several enlightening discussions. The Westerbork Synthesis
Radio Telescope is operated by ASTRON (The Netherlands Institute for
Radio Astronomy) with support from the Netherlands Foundation for
Scientific Research (NWO). The Digitized Sky Surveys were produced at
the Space Telescope Science Institute under U.S. Government grant
NAG W-2166. The images of these surveys are based on photographic
data obtained using the Oschin Schmidt Telescope on Palomar Mountain
and the UK Schmidt Telescope. The plates were processed into the
present compressed digital form with the permission of these
institutions. We thank the anonymous referee for carefully reading
the manuscript, and providing helpful comments which have improved
the paper.
\end{acknowledgements}

\bibliographystyle{aa}
\bibliography{sings}

\appendix

\section{Deconvolution of the Faraday dispersion function\label{appendix:rmclean}}

The Faraday dispersion function recovered using RM-Synthesis has
undesirable features resulting from the incomplete sampling in the
$\lambda^2$ domain. These effects can be reduced using a deconvolution
procedure, similar to what is done in the synthesis imaging
case. Here, we outline and justify our procedure in greater detail
than was provided in \S\,\ref{subsection:rmclean}.

There are two salient features of the RMSF: the width of the main lobe
($\Delta\phi$), and the sidelobe structure. In the WSRT-SINGS survey,
the sampling function (see Figure \ref{figure:RMSFs}, panel {\em e})
can be described as two windows. The form of our RMSF can be likened
to the interference pattern in a double-slit experiment. If the
windows (slits) were infinitely narrow, the sidelobes would have unit
amplitude and the spacing would be inversely related to the distance
in $\lambda^2$ between the windows. In fact, the locations on the
$\phi$-axis of the sidelobes would be identical to the $n\pi$
ambiguity which would result from a standard rotation measure
determination using the two $\lambda^2$ frequency samples alone. In
actuality, the windows are made up of many individual frequency
measurements, and therefore each effectively have a finite width.
This provides a
taper to damp down the sidelobes. But the windows are still relatively
narrow compared to the distance between them, so the sidelobes remain
rather high. It is therefore desirable to remove these sidelobes from
the recovered Faraday dispersion function, particularly in cases which
may potentially contain multiple distinct sources or broadening along
the $\phi$ axis. Before deconvolution, the sidelobe structure of a
brighter component contaminates the response to fainter components,
and may shift the location of peaks relative to their true position.

The mechanism for performing this deconvolution is quite similar to
the {\tt CLEAN} algorithm which is already well known from the
synthesis imaging case. We first describe the algorithm. Then we move
on to justifying our choice of working in a domain in which
$\lambda_0^2\,\neq\,0$ (see equation \ref{equation:phiinvsum}).
Here we demonstrate that deconvolution in that
space is equivalent to deconvolution in the ``true'' measurement
space. Next we describe how a restoring RMSF was selected, and
conclude with some general remarks about the routine.

\subsection{The deconvolution procedure}

The algorithm, also described by \citet{brentjens_2007}, consists of
the following steps:
\begin{enumerate}
\item In each spatial pixel, the complex ($Q(\phi)$,$U(\phi)$)
  spectrum is cross-correlated with the complex RMSF. The location of
  the peak absolute value of the cross-correlation, $\phi_m$, is
  noted.
\item If $P(\phi_m)$ is greater than a user-defined cutoff, a shifted
  and scaled version of the complex RMSF is subtracted from the
  complex ($Q(\phi)$,$U(\phi)$) spectrum. The scaled RMSF is
  $gP(\phi_m)R(\phi-\phi_m)$, where $g$ is a (real) gain factor.
\item The value $gP(\phi_m)$ is stored as a ``clean component''.
\item Steps 1--3 are repeated until the value of $P(\phi_m)$ is no
  longer higher than the cutoff, or a maximum number of iterations
  have been performed.
\item Finally, the clean components are convolved with a restoring
  Gaussian beam with a FWHM equal to $2\sqrt{3}/\Delta\lambda^2$ (see
  \ref{subappendix:restor}), and added to the residual $F(\phi)$. The
  result is the deconvolved $F(\phi)$ spectrum.
\end{enumerate}
The reason for using a cross-correlation in step (1), rather than
simply searching for the peak $P(\phi)$ (in analogy to imaging
deconvolution), is the hope that incorrect component localization due
to possible sidelobe confusion will be eliminated. In practice, the
two techniques were found to yield the same results in our data.

\subsection{Deconvolution in the ``shifted'' domain}

We begin with the definitions of the dirty Faraday dispersion function
in its unshifted (pure) form:
\begin{equation}
\tilde{F}(\phi)\,=\,F(\phi)\,\ast\,R(\phi)\,=\,K\,\int_{-\infty}^{+\infty}\,\tilde{P}(\lambda^2)e^{-2i\phi\lambda^2}\,d\lambda^2
\end{equation}
and its shifted form:
\begin{equation}
\tilde{F}'(\phi)\,=\,K\,\int_{-\infty}^{+\infty}\,\tilde{P}(\lambda^2)e^{-2i\phi(\lambda^2-\lambda_0^2)}\,d\lambda^2
\end{equation}
\begin{equation}\tilde{F}'(\phi)\,=\,K\,\int_{-\infty}^{+\infty}\,\tilde{P}(\lambda^2)e^{-2i\phi\lambda^2}\,e^{+2i\phi\lambda_0^2}\,d\lambda^2
\end{equation}
Noting that the factor $e^{2i\phi\lambda_0^2}$ is constant in $\lambda^2$, we can pull it out of the integral as a constant factor.
\begin{equation}
\tilde{F}'(\phi)\,=\,K\,e^{2i\phi\lambda_0^2}\,\int_{-\infty}^{+\infty}\,\tilde{P}(\lambda^2)e^{-2i\phi\lambda^2}\,d\lambda^2,
\end{equation}
which is seen to be equivalent to:
\begin{equation}
\tilde{F}'(\phi)\,=\,e^{2i\phi\lambda_0^2}\,\tilde{F}(\phi).
\end{equation}
In exactly the same way, it can be shown that the same relation holds for the RMSF:
\begin{equation}
R'(\phi)\,=\,e^{2i\phi\lambda_0^2}\,R(\phi).
\end{equation}
We also must define
\begin{equation}
F'(\phi)\,\equiv\,e^{2i\phi\lambda_0^2}\,F(\phi),
\end{equation}
which is just the {\em actual} Faraday dispersion function multiplied by the ``shift'' factor.

Again beginning with the shifted, dirty Faraday dispersion function,
\begin{equation}
\tilde{F}'(\phi)\,=\,e^{2i\phi\lambda_0^2}\,\tilde{F}(\phi)\,=\,e^{2i\phi\lambda_0^2}\,\times\,\left[F(\phi)\,\ast\,R(\phi)\right]
\end{equation}
and incorporating the definition of a convolution,
\begin{equation}
\tilde{F}'(\phi)\,=\,e^{2i\phi\lambda_0^2}\,\int_{-\infty}^{+\infty}\,F(\phi-u)\,R(u)\,du.
\end{equation}
We can now move the exponential inside the integral, and introduce two exponentials whose product is unity:
\begin{equation}
\tilde{F}'(\phi)\,=\,\int_{-\infty}^{+\infty}\,\left(e^{2iu\lambda_0^2}\,e^{-2iu\lambda_0^2}\right)\,e^{2i\phi\lambda_0^2}\,F(\phi-u)\,R(u)\,du.
\end{equation}
Rearranging,
\begin{equation}
\tilde{F}'(\phi)\,=\,\int_{-\infty}^{+\infty}\,\left(F(\phi-u)e^{2i(\phi-u)\lambda_0^2}\right)\,\left(R(u)e^{2iu\lambda_0^2}\right)\,du
\end{equation}
\begin{equation}\tilde{F}'(\phi)\,=\,\int_{-\infty}^{+\infty}\,F'(\phi-u)\,R'(u)\,du
\end{equation}
which is just
\begin{equation}
\tilde{F}'(\phi)\,=\,F'(\phi)\,\ast\,R'(\phi).
\end{equation}
This means that the shifted, dirty Faraday dispersion function is the
convolution of the shifted RMSF $R'(\phi)$ with the ``rolled-up''
Faraday dispersion function $F'(\phi)$. After deconvolution, then, we
will have determined $F'(\phi)$. Recovery of the actual goal,
$F(\phi)$, can be realized by multiplying by the ``inverse shift factor,''
$e^{-2i\phi\lambda_0^2}$.

\subsection{Selection of a restoring beam\label{subappendix:restor}}

The width of the main lobe of the RMSF, $\Delta\phi$, reflects the
fact that we have only limited precision in our determination of the
Faraday depth of a polarized source. This is fundamentally related to
the sampling function. Thus, once we have extracted point source
components during the deconvolution routine, they must be replaced at
a resolution appropriate to the measured frequency domain. In other
words the point source model is convolved with a ``restoring
beam''. How does one choose the form of this restoring function?

By derotating to $\lambda_0^2\,\neq\,0$,
\citet{brentjens_debruyn_2005} show that the variation in the
imaginary part of the RMSF is minimized. In a sense, we have selected
a frame which rotates as a function of $\lambda^2$ in $(Q,U)$ space
such that the polarization vector stays along the $Q$ axis as much as
possible. We would like to choose a restoring function which is
equivalent to zeroing out the residual $U$ response in this rotating
frame: in other words, asserting that we have in fact determined the
true rotation measure of the polarized source. (Recall, this is
equivalent to eliminating the sidelobes of the RMSF.) Thus, we set the
imaginary part of the restoring function to zero, and the real part to
a Gaussian similar to the central lobe of the real part of the RMSF,
{\em in the rotating frame} $\lambda^2-\lambda_0^2$. Formally, we
choose
\begin{equation}
R(\phi)\,=\,e^{-\phi^2/2\sigma^2},
\label{eqn:restor}
\end{equation}
a real-valued function, where $\sigma$ is selected to match the width
of the real part of the shifted RMSF. As in the case of image
plane deconvolution, it has proven to be useful to retain a main
lobe width that is matched to the ``dirty'' image. While a narrower
restoring RMSF might be selected, this would result in a mismatched
resolution when compared to the deconvolution residuals.
After the restoring function in eqn. \ref{eqn:restor}
is used to smooth the point source model to the appropriate resolution,
it is put back into the measurement domain by multiplying with the
inverse shift factor.

\subsection{Practical considerations}

Formally, this technique is straightforward. In the presence of noise,
however, one must take care to properly treat the residuals and the
{\tt CLEAN} components which are returned by the deconvolution
algorithm. In practice, the residuals are multiplied by the inverse
shift factor (described above) separately. Then, the
model is convolved with the restoring beam selected above, multiplied by
the inverse shift factor, and added to the residuals. The final result is
our deconvolved Faraday dispersion function, as used in this paper.

A final consideration is that the cubes should be reordered prior to
performing the deconvolution routine. This is because the routine
works not on individual image slices, but rather on spectra along the
$\phi$ axis. Therefore it is considerably more efficient for the
routine to be able to read out spectra sequentially, instead of having
to read through the full cube to construct each spectrum.

This {\tt RM-CLEAN} algorithm has been implemented by us as a {\tt
  MIRIAD} task and has been made publically available\footnote{The
  {\tt RM-CLEAN} procedure is available for download from {\tt
    <http://www.astron.nl/$\mathtt{\sim}$heald/software>}.}. The task
typically takes less than 20 minutes to operate on a pair of
($Q(\phi)$,$U(\phi)$) cubes with dimensions
($512\,\times\,512\,\times\,401$) on a dual 1.8\,GHz Opteron system,
to the cutoff level described in Section \ref{subsection:rmclean}.

\end{document}